\date{}
\documentclass[preprint,prd,12pt]{revtex4-1}
\usepackage{amsmath}
\usepackage{amssymb}
\usepackage{mathrsfs} 
\usepackage{graphicx} \usepackage{bm} \usepackage{epsfig}
\usepackage{epstopdf} 
\DeclareGraphicsExtensions{.pdf,.eps,.png,.jpg,.mps} 
\usepackage[centerlast]{caption}
\usepackage{subcaption}
\usepackage[english]{babel}
\def\Tr{\mathop{\rm Tr}}
\usepackage{color}

\begin{document}

\title{Hartree approximation in curved spacetimes revisited II: The semiclassical Einstein equations and de Sitter self-consistent solutions}

\author{Diana L. L\'opez Nacir $^{1,2}$}
\author{Francisco D. Mazzitelli$^{3}$}
\author{Leonardo G. Trombetta$^2$}

\affiliation{$^1$ Abdus Salam International Centre for Theoretical Physics
Strada Costiera 11, 34151, Trieste, Italy}
\affiliation{$^2$ Departamento de F\'\i sica and IFIBA, FCEyN UBA, Facultad de Ciencias Exactas y Naturales,
 Ciudad Universitaria, Pabell\' on I, 1428 Buenos Aires, Argentina}
\affiliation{$^3$ Centro At\'omico Bariloche 
Comisi\'on Nacional de Energ\'\i a At\'omica,
R8402AGP Bariloche, Argentina}

\date{\today}

\begin{abstract}
  We consider the semiclassical Einstein equations (SEE) in the presence of a quantum scalar field with self-interaction $\lambda\phi^4$. Working  in  the  Hartree truncation of the two-particle irreducible (2PI) effective action, we compute the vacuum expectation value of the energy-momentum tensor of the scalar field, which act as a source of the SEE. We obtain the renormalized SEE by implementing a consistent 
 renormalization procedure. We apply our results to find self-consistent de Sitter solutions to the SEE in situations with or without spontaneous breaking of the $Z_2$-symmetry.

\end{abstract}


\maketitle

\section{Introduction}

Quantum field theory in curved spacetimes  \cite{birrell,wald,fulling,parker} is the natural framework
for the study of quantum phenomena in situations where the  gravitation itself can be treated classically. 
 Of special interest is  quantum field theory in de Sitter spacetime.
In fact, de Sitter  spacetime plays a central role  in most of inflationary models of  the early Universe \cite{Inflation2, Inflation3, Inflation4}, where the
energy density and pressure of the inflaton field act approximately as a cosmological constant.
Moreover,  the amplification  of quantum fluctuation during an inflationary period  with  an approximately  de Sitter background metric,    gives a natural mechanism for generating 
 nearly scale-invariant spectrum of primordial inhomogeneities, which can successfully explain the observed CMB anisotropies \cite{Inflationandacc1,Inflation1}.
De Sitter  spacetime  is also potentially important for  understanding the final fate of the Universe if the current
accelerated expansion is  due to a small cosmological constant, which nowadays is a possibility that  is compatible with  observations \cite{Inflationandacc1,Acc1,Acc2,Acc3}.
On the other hand, previous studies of interacting quantum scalar fields in de Sitter spacetime have revealed 
  that  the standard perturbative expansion  gives rise to  corrections  that secularly grow with time and/or infrared divergences  \cite{Weinberg, Starobinski,Woodard,Meulen,Seery1,SeeryR,TanakaR,Shandera},  signaling a possible  deficiency of the perturbative approach. 
This  has motivated several authors to consider  alternative techniques (see for instance  \cite{Starobinski,Woodard,ProkopecN,Rajaraman,Hollands,Beneke,Shandera,Riotto,Boyanovsky,Akhmedov}) and in particular, to use nonperturbative resummation schemes 
\cite{Rigopoulos, Serreau, Serreau0,Youssef,Arai, Rigo2,Nos1}.

In the above situations, it is important  to study not only test fields evolving on a fixed background, but also
 to  take into account the backreaction of the quantum fields on the dynamics of the spacetime geometry.
The backreaction problem has been  explored by a number of authors in the context of semiclassical gravity (see for instance \cite{BR1,BR2,BR3,Simon,Flanagan}),
where
the dynamics of the classical metric is governed by the so-called Semiclasical Einstein Equations (SEE). The SEE are a generalization 
of the Einstein equations that contain as a source the expectation value of the energy-momentum tensor of the quantum matter fields,  $\langle T_{\mu\nu}\rangle$ \cite{birrell,wald, fulling,parker}. Self-consistent de Sitter solutions have been found for the case of free quantum fields  \cite{Starobinski2,Vilenkin,Castagnino,DiegDi,wada-azuma}.
The influence of the initial state of the quantum field on the semiclassical solutions has been studied in Refs. \cite{Anderson,PerezNadal}.

Since   $\langle T_{\mu\nu}\rangle$  is formally a divergent quantity, in order to address the backreaction problem it is necessary to analyze  the renormalization process. 
For free and interacting quantum fields in the one-loop approximation, there are   well known covariant renormalization methods   \cite{birrell,wald,fulling,parker}.
Our main goal in this work is to improve the current understanding of these  methods in the case in which the quantum effects are taken into account 
nonperturbatively.  For this,  we consider a quantum self-interacting scalar field in the Hartree approximation,
which corresponds to the simplest nonperturbative truncation to the two-particle irreducible effective action (2PI EA), introduced  by Cornwall, Jackiw and Tomboulis \cite{CJT}.
The Hartree (or Gaussian) approximation involves the resummation of a particular type of Feynman diagrams which are called superdaisy (see for instance \cite{Amelino}) to an infinite perturbative order. 
This approximation can also be introduced  by means of a variational principle \cite{Stevenson, autonomous}. However,  the use of the 2PI EA  is advantageous for at least two reasons.
First, it  provides a framework  for resumming classes of diagrams that can be  systematically improved.
Second, for any truncation of the EA, it implies certain consistency relations between different counterterms that  allow a renormalization procedure that is consistent with the standard  perturbative (loop-by-loop) renormalization of the bare coupling constants \cite{Bergesetal}. The latter is crucial for the consistent renormalization procedure developed in Ref. \cite{Bergesetal} for  Minkowski spacetime, which  in \cite{Nos1}  (from now on paper I), using the same model considered here,  we have extended to general curved background metrics.

The renormalization problem of the SEE in the Hartree approximation has been considered previously in \cite{mazzi-paz,Arai}. However, 
it has not been analyzed using the consistent  renormalization procedure \cite{Bergesetal} that   we extended to curved spacetimes in paper I
in order to  renormalize the field and gap equations.
Our focus in this paper is to prove that the same set
of renormalized parameters  leads to  SEE that can be made finite, and independent on the arbitrary scale introduced by the regularization scheme
(which for the field and gap equations was explicitly shown in paper I), by  suitable renormalizations of the bare gravitational constants.

The paper is organized as follows. In Sec.  \ref{2PISect} we introduce the 2PI EA in curved spacetimes. In Sec. \ref{P1Sect}  we  present our model and summarize the main relevant results of   paper I  for the renormalization of the mass and coupling constant of the field.  The reader acquainted with  paper I may skip this section.  In Sec. \ref{SEESect} we show that the same counterterms that make finite the field and gap equations can also be used to absorb the non-geometric divergences in the SEE, extending  the consistent renormalization procedure to the gravitational sector. The geometric divergences  can be absorbed into the usual gravitational counterterms.
In Sec. \ref{DSSect} we analyze the field, gap and SEE in  de Sitter spacetimes. The high  symmetry of these spacetimes allows us to compute explicitly the two point function and the energy-momentum tensor, to end with a set of algebraic equations that determine self-consistently  the mean value of the field and the de Sitter curvature. We will present some numerical solutions to these equations.
In Sec. \ref{ConcluSect} we include our conclusions.
Throughout the paper we set $c=\hbar=1$ and adopt the mostly plus sign convention.

\section{The 2PI effective action}\label{2PISect}

A detailed description to the 2PI EA formalism can be found in several papers and textbooks, such as \cite{calzetta,CJT,Ramsey}. 
 In  this section, in order to  make this work as self-contained as possible and to set the notation, we briefly  summarize 
the main relevant aspects of the formalism applied to a self-interacting scalar field $\phi$ in a general curved spacetime. 
 
The 2PI generating functional can be written as \cite{Bergesetal} 
\begin{equation}
 \Gamma_{2PI}[\phi_0,G,g^{\mu \nu}] = S_0[\phi_0,g^{\mu \nu}] + \frac{i}{2} \textup{Tr} \ln(G^{-1}) + \frac{i}{2} \textup{Tr}( G_0^{-1} G) + \Gamma_{int}[\phi_0,G,g^{\mu \nu}],
\label{gamma-2PI-alt}
\end{equation}
where $S_0$ is quadratic part of the classical action $S$ without any counterterms,
\begin{equation}
i G_0^{ab}(x,x') = \frac{1}{\sqrt{-g}} \frac{\delta^2 S_0[\phi_0,g^{\mu\nu}]}{\delta \phi_a(x) \delta \phi_b(x')} \frac{1}{\sqrt{-g'}},
\end{equation} and
\begin{equation}
 \Gamma_{int}[\phi_0,G,g^{\mu \nu}] = S_{int}[\phi_0,g^{\mu \nu}] + \frac{1}{2} \Tr \left[ \frac{\delta^2 S_{int}}{\delta \phi_0 \delta \phi_0} G \right] + \Gamma_2[\phi_0,G,g^{\mu \nu}],
\end{equation} where the functional $\Gamma_2$ is $-i$ times the sum of all two-particle-irreducible vacuum-to-vacuum diagrams with lines given by $G$ and vertices obtained from the shifted action $S^{F}_{int}$, which comes from expanding $S_{int}[\phi_0 + \varphi]$ and collecting all terms higher than quadratic in the fluctuating field $\varphi$.
Here $a, b$ are time branch indices (with index set $\{+,-\}$ in the usual notation) corresponding to the ordering on the contour in  the  ``closed-time-path''(CTP) or  Schwinger-Keldysh  \cite{calzetta} formalism.

The equations of motion for the field and propagator are obtained by 
\begin{subequations}
\begin{align}
 \frac{\delta \Gamma_{2PI}}{\delta \phi_0}\Big{|}_{\phi_{+}=\phi_{-}=\phi; g^{\mu\nu}_{+}=g^{\mu\nu}_{-}=g^{\mu\nu}} &= 0,\\
 \frac{\delta \Gamma_{2PI}}{\delta G}\Big{|}_{\phi_{+}=\phi_{-}=\phi;g^{\mu\nu}_{+}=g^{\mu\nu}_{-}=g^{\mu\nu}}  &= 0.
\end{align}
\end{subequations}
To arrive at the SEE we extremize the combination $S_g[g^{\mu\nu}] + \Gamma_{2PI}[\phi_0,G,g^{\mu\nu}]$ with respect to the metric,
\begin{equation} \label{SEEdef}
 \frac{\delta\left(S_g[g^{\mu\nu}] + \Gamma_{2PI}[\phi_0,G,g^{\mu\nu}]\right)}{\delta g^{\mu\nu}} \Big{|}_{\phi_{+}=\phi_{-}=\phi;g^{\mu\nu}_{+}=g^{\mu\nu}_{-}=g^{\mu\nu}} = 0,
 \end{equation}
where $S_g$ is the gravitational action. As it is well known \cite{birrell,wald,fulling}, this equation is formally divergent, with the divergences contained in the 
 vacuum expectation value of the energy-momentum tensor $\langle T_{\mu\nu}\rangle$, defined by
\begin{equation}\label{tmunu-def}
\langle T_{\mu\nu}\rangle=-\frac{2}{\sqrt{-g}}\frac{\delta\Gamma_{2PI}[\phi_0,G,g^{\mu\nu}]}{\delta g^{\mu\nu}} \Big{|}_{\phi_{+}=\phi_{-}=\phi;g^{\mu\nu}_{+}=g^{\mu\nu}_{-}=g^{\mu\nu}}.
\end{equation} It is also well known \cite{birrell,wald,fulling} that the renormalization procedure requires the inclusion of terms quadratic in the curvature  in the gravitational action, so that
\begin{equation}\label{Sg}
S_{g}=\frac{1}{2}\int d^4 x \sqrt{-g}\left\{
\kappa_B^{-1}(R-2\Lambda_B)-\alpha_{1B} R^2-\alpha_{2B}
R_{\mu\nu}R^{\mu\nu}-\alpha_{3B}
R_{\mu\nu\rho\sigma}R^{\mu\nu\rho\sigma}\right\},
\end{equation} where
$R_{\mu\nu\rho\sigma}$ is the curvature tensor,
$R_{\mu\nu}=R^{\rho}_{\mu\rho\nu}$, and $\kappa_B=8\pi G_N^B$, $\Lambda_B$, 
$\alpha_{iB}$ ($i=1,2,3$) are bare parameters which are to be appropriately chosen to cancel the divergences in $\langle T_{\mu\nu}\rangle$.

\section{$\lambda \phi^4$ theory in the Hartree approximation: renormalization of the field and gap equations}\label{P1Sect}

We consider a nonminimally coupled scalar field with quartic self-coupling in a curved background with metric $g_{\mu \nu}$. The corresponding classical action reads
\begin{equation}
 S_m[\phi,g^{\mu\nu}] = -\int d^4 x \, \sqrt{-g} \left[ \frac{1}{2} \phi \left( -\square + m^2_B + \xi_B R \right) \phi + \frac{1}{4!} \lambda_B \phi^4 \right],
 \label{classical-S}
\end{equation}
where $\square = \frac{1}{\sqrt{-g}} \partial_\mu \left( \sqrt{-g} g^{\mu \nu} \partial_\nu \right)$, $g \equiv \det (g_{\mu\nu})$. 
\begin{figure}
\centering
\includegraphics[scale=0.175]{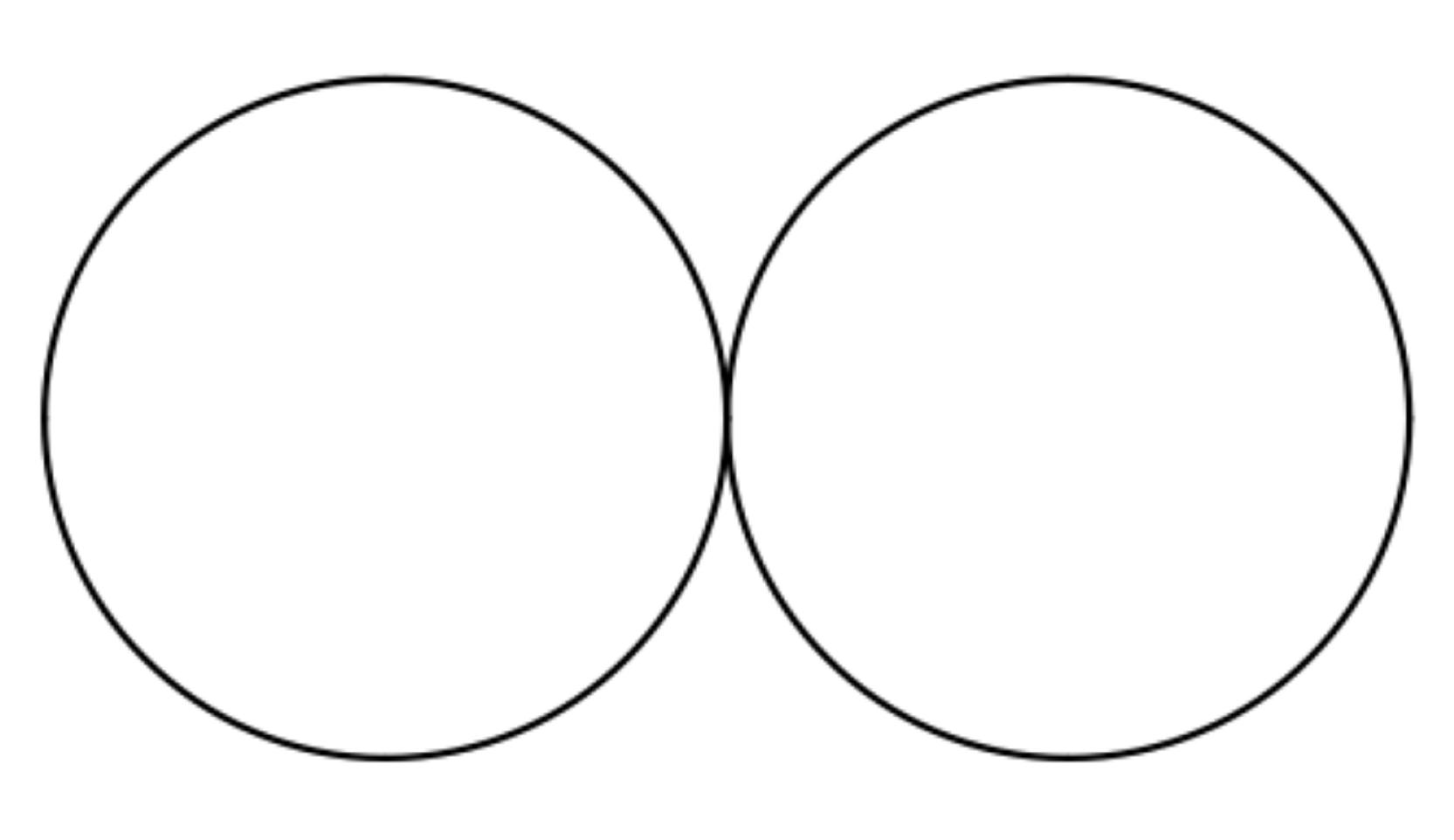}
\vspace{0.5cm}
\caption{ 2PI  ``double-bubble'' diagram .}\label{diagrams}
\end{figure}
In the  Hartree approximation, which corresponds to the inclusion of  only the double-bubble diagram shown in Fig. \ref{diagrams}, the
 2PI effective action is given by
\begin{eqnarray}
 \Gamma_{2PI}[\phi_0,G,g_{\mu\nu}] &=& -\int d^4 x \, \sqrt{-g} \left[ \frac{1}{2} \phi_0 \left( -\square + m_{B2}^2 + \xi_{B2} R  \right) \phi_0 + \frac{1}{4!} \lambda_{B4} \phi_0^4 \right] + \frac{i}{2} \Tr \ln ( G^{-1} ) \nonumber \\ 
&&- \frac{1}{2} \int d^4 x \, \sqrt{-g} \left[ -\square + m_{B0}^2 + \xi_{B0} R + \frac{1}{2} \lambda_{B2} \phi_0^2 \right] G(x,x) \label{2PI-lambdaphi4} \\
\nonumber
&& - \frac{\lambda_{B0}}{8} \int d^4 x \, \sqrt{-g} \, G^2(x,x),  
\end{eqnarray}where, for the sake of simplicity,  we drop the time branch indices, since for the Hartree approximation it is known that the CTP formalism gives the same equations of motion than the usual 
{\it in-out} formalism  \cite{Ramsey}. 

Taking the variation with respect to $\phi_0$ and $G$ 
we obtain equations of motion for the mean field and the propagator:
\begin{eqnarray}
 \left( -\square + m_{B2}^2 + \xi_{B2} R + \frac{\lambda_{B4}}{6}  \phi_0^2 + \frac{\lambda_{B2}}{2}  [G] \right) \phi_0(x) &=& 0, \label{unrenfield}\\
 \left( -\square + m^2_{B0} + \xi_{B0} R + \frac{\lambda_{B2}}{2}  \phi_0^2 + \frac{\lambda_{B0}}{2}  [G] \right) G(x,x') &=& -i \frac{\delta(x-x')}{\sqrt{-g'}},\label{unrenprop}
\end{eqnarray}with $[G]$ the coincidence limit of the propagator $G(x,x')$.

It is important to note that here we are  taking into account the possibility of   having  different counterterms  for a given  parameter of the classical  action Eq. (\ref{classical-S}). These are  
denoted  using  different  subscripts in the bare parameters that refer to the power of $\phi_0$ in the corresponding term of the action.
In  the  Hartree approximation, this point turns out to  be  crucial  for the implementation of the consistent renormalization procedure described in \cite{Bergesetal} .  
  Indeed, as  shown in  \cite{Bergesetal}
(see also  Appendix A of paper I),  there are various possible $n$-point functions that can  be obtained from functionally differentiating $\Gamma_{2PI}[\phi_0,G,g^{\mu\nu}]$ with 
respect to $\phi_a$ and $G_{ab}$, which in the exact theory must satisfy certain consistency conditions.  On the other hand, for any truncation of the 2PI EA, the validity of such
consistency conditions is not guarantee. However,    one can find a relation between  the  different counterterms   by imposing  the consistency conditions 
at a given renormalization point. Doing this, any possible deviation of the consistency conditions is finite and under perturbative control. 
In other words, had we not allowed for different counterterms, the diagrams contributing to the consistency
conditions could contain perturbative divergent contributions which could not be absorbed anywhere.

In our case, the consistency conditions for the two- and four-point functions,  evaluated at  $\phi_0=0$,  are given by

\begin{equation}
  \frac{\delta^2 \Gamma_{int}}{\delta \phi_1 \delta \phi_2} \Bigg|_{\phi=0} = 2 \frac{\delta \Gamma_{int}}{\delta G_{12}} \Bigg|_{\phi=0},
\label{2-pt-relation}
\end{equation}
and 
\begin{eqnarray}
\frac{\delta^4 \Gamma_{1PI}[\phi_0]}{\delta \phi_1 \delta \phi_2 \delta \phi_3 \delta \phi_4}\Bigg|_{\phi_0=0} = 2 \left[ \frac{\delta^2 \Gamma_{int}}{\delta G_{12} \delta G_{34}} \Bigg|_{\bar{G},\phi_0=0} + perms(2,3,4) \right] - \frac{1}{2} \frac{\delta^4 \Gamma_{int}}{\delta \phi_1 \delta \phi_2 \delta \phi_3 \delta \phi_4} \Bigg|_{\bar{G},\phi_0=0}, \,\,\,\,\,\,
 \label{4-pt-relation}
\end{eqnarray}
where
\begin{equation}
\Gamma_{1PI}[\phi_0,g^{\mu \nu}] = \Gamma_{2PI}[\phi_0,\bar{G}[\phi_0],g^{\mu \nu}].
 \label{1PI-2PI}
\end{equation}

In what follows  we consider two different parametrizations of the bare couplings:
\begin{subequations}
\begin{align}
 m^2_{Bi} &= m^2 + \delta m_i^2 = m_R^2 + \delta \tilde{m}_i^2\,\,(i=0,2), \\
 \xi_{Bi} &= \xi + \delta \xi_i = \xi_R + \delta \tilde{\xi}_i \,\,(i=0,2),\\
 \lambda_{Bi} &= \lambda + \delta \lambda_i = \lambda_R + \delta \tilde{\lambda}_i, \,\,(i=0,2,4).
\end{align}
\label{type-of-counterterms}
\end{subequations}
The first separation corresponds to the MS scheme (i.e., the counterterms $\delta m_i^2$, $\delta \xi_i$ and $\delta \lambda_j$ ($i=0,2$,$j=0,2,4$) contain
only divergences and no finite part), 
while in the second separation $m_R^2$, $\xi_R$ and $\lambda_R$ are chosen 
to be the renormalized parameters as defined from the effective potential (see below).

By imposing the conditions (\ref{2-pt-relation}) and (\ref{4-pt-relation}), one can obtain the  following relation
between the different counterterms \cite{Nos1}:
\begin{subequations}\label{relcounterterms}
\begin{align}
  &\delta m_0^2 = \delta m_2^2 \equiv \delta m^2, \label{masses} \\
  &\delta \xi_0 = \delta \xi_2 \equiv \delta \xi, \label{xis} \\
  &\delta \lambda_0= \delta \lambda_2, \label{lambdas0-2}\\
 &\delta \lambda_4 - 3 \delta \lambda_2 = 2( \lambda - \lambda_R),
 \label{lambdas2-4}
\end{align}
\end{subequations}
 with
\begin{equation}
 \frac{\delta^4 \Gamma_{1PI}[\phi_0]}{\delta \phi_1 \delta \phi_2 \delta \phi_3 \delta \phi_4} \Bigg|_{\phi_0=0} = - \lambda_R \delta_{12} \delta_{13} \delta_{14},\label{der1PI}
\end{equation} where
we used $\phi_i \equiv \phi_0(x_i)$ as a notational shorthand. Recalling that  the effective potential
 is proportional to the effective action at a constant value of $\phi_0$,  the renormalized self-interaction coupling $\lambda_R$
  can be also written as
\begin{equation}
\lambda_R = \frac{d^4 V_{eff}}{d \phi_0^4} \Bigg|_0. 
\end{equation}

With the use of these relations, one can recast Eqs. (\ref{unrenfield}) and (\ref{unrenprop})   as
\begin{eqnarray}
 \left( -\square + m_{ph}^2 + \xi_R R - \frac{1}{3} \lambda_R \phi_0^2 \right) \phi_0(x) &=&0, \label{field-eq} \\
 \left( -\square + m_{ph}^2 + \xi_R R \right) G_1(x,x') &=& 0,\label{gap-eq}
 \end{eqnarray} where $m_{ph}^2$ is identified with the physical mass of the fluctuations and satisfies a self-consistent equation (i.e., the gap equation) that reads
 \begin{equation}
 m_{ph}^2 + \xi_R R = m^2 + \delta m^2+ (\xi + \delta \xi) R + \frac{1}{2} (\lambda + \delta \lambda_2) \phi_0^2 + \frac{1}{4} (\lambda + \delta \lambda_2) [G_1].
 \label{phys-mass}
\end{equation}

A point that is worth emphasizing  here is that these relations cannot be imposed in an arbitrary spacetime metric, since the renormalized
parameters must be constant, while the fourth derivative of 1PI EA in Eq. (\ref{der1PI}) might not. However, in order to define the renormalized parameters,
one can choose  a particular fixed background metric with constant curvature invariants as the renormalization point at which the consistency conditions are imposed. 
In paper I we  considered both  Minkowski and de Sitter spacetimes. Here, for the sake of generality, we will also consider both renormalization points. Therefore, 
we define the renormalized parameters as those derived from the effective potential and evaluated for a fixed de Sitter spacetime with $R=R_0$,
\begin{subequations}
\begin{align}
\mathcal{M}_R^2 &\equiv \frac{d^2 V_{eff}}{d \phi_0^2} \Bigg|_{\phi_0=0,R=R_0} = \mathcal{M}_{ph}^2(\phi_0=0,R=R_0), \label{ren-mass-def-dSren}  \\
\xi_R &\equiv \frac{d^3 V_{eff}}{d R \, d \phi_0^2} \Bigg|_{\phi_0=0,R=R_0} = \frac{d \mathcal{M}_{ph}^2}{d R} \Bigg|_{\phi_0=0,R=R_0}, \label{ren-xi-def-dSren} \\
\lambda_R &\equiv \frac{d^4 V_{eff}}{d \phi_0^4} \Bigg|_{\phi_0=0,R=R_0} = 3 \frac{d^2 \mathcal{M}_{ph}^2}{d \phi_0^2} \Bigg|_{\phi_0=0,R=R_0} - 2 \lambda_R, \label{ren-lambda-def-dSren}
\end{align}
\label{ren-quantities-dSren}
\end{subequations} where we are using the notation $\mathcal{M}_R^2=m_R^2+\xi_R R$.  In  particular, the limit $R_0\to 0$ could be taken to recover
the usual renormalized parameters defined  in  Minkowski spacetime.  

In order to obtain the renormalized  gap equation it is useful to consider the adiabatic expansion of the 
propagator at the coincidence limit:
 \begin{eqnarray}
 [G_1] &=& \frac{1}{8 \pi^2} \left( \frac{m_{ph}^2}{\mu^2} \right)^{\epsilon/2} \sum_{j \geq 0} \, [\Omega_j] (m_{ph}^2)^{1-j} \, \Gamma \left( j-1 - \frac{\epsilon}{2} \right) \notag\\
  &\equiv& \frac{1}{4\pi^2 \epsilon} \left[ m_{ph}^2 + \left(\xi_R - \frac{1}{6} \right) R \right]  + 2\, T_F (m_{ph}^2,\xi_R,R,\tilde{\mu}), \label{adiab-G1} 
\end{eqnarray} where $\epsilon = n - 4$, $\Gamma(x)$ is the Gamma function, and the Schwinger-DeWitt coefficients $[\Omega_j]$ 
are scalars of adiabatic order $2j$
built from the metric and its derivatives and satisfy certain recurrence relations.
In the second line, we have used the explicit expressions for the  coefficients $[\Omega_0]=1$ and $[\Omega_1]=-(\xi_R-1/6)R$, given in \cite{mazzi-paz2},  we have expanded for $\epsilon \to 0$ and we have
redefined $\mu \to \tilde{\mu}$ to absorb some constant terms, defining
\begin{eqnarray}
\nonumber T_F (m_{ph}^2,\xi_R,R,\tilde{\mu})& =& \frac{1}{16 \pi^2}  \Bigg{\{} \left[ m_{ph}^2 + \left(\xi_R - \frac{1}{6} \right) R \right] \ln \left( \frac{m_{ph}^2}{\tilde{\mu}^2} \right) + \left(\xi_R - \frac{1}{6} \right) R \\ 
&- & 2 F(m_{ph}^2,\{R\}) \Bigg{\}},
\end{eqnarray} where the function $F(m_{ph}^2,\{R\})$ contains the adiabatic orders higher than two, is independent of $\epsilon$ and $\mu$, and satisfies the following properties:
\begin{subequations}
\begin{align}
&F(m_{ph}^2,\{R\})\bigg|_{R_{\mu\nu\rho\sigma} = 0} = 0, \label{F-prop1} \\
&\frac{dF(m_{ph}^2,\{R\})}{d m_{ph}^2} \Bigg|_{R_{\mu\nu\rho\sigma}=0} = 0,\label{F-prop2} \\
&\frac{dF(m_{ph}^2,\{R\})}{d R} \Bigg|_{R_{\mu\nu\rho\sigma}=0,\phi_0=0} = 0. \label{F-prop3}
\end{align}
\label{F-props}
\end{subequations}

Taking into account the relations in Eq. (\ref{relcounterterms})  between the counterterms,  the gap equation can be made finite
with the use of the following MS counterterms:
\begin{subequations}
\begin{align}
\delta m^2 &= -\frac{\lambda }{16\pi^2 \epsilon} \frac{m^2}{1 + \frac{\lambda }{16\pi^2 \epsilon}}, \\
\delta \xi &= -\frac{\lambda }{16\pi^2 \epsilon} \frac{\left( \xi - \frac{1}{6} \right)}{1 + \frac{\lambda}{16\pi^2 \epsilon}}, \\
\delta \lambda_2 &= -\frac{\lambda }{16\pi^2 \epsilon} \frac{\lambda}{1 + \frac{\lambda }{16\pi^2 \epsilon}}. 
\end{align}
\end{subequations} Once made finite and written in terms of the MS parameters, it reads
\begin{eqnarray}
 m_{ph}^2 + \xi_R R &=& m^2 + \xi R + \frac{1}{2} \lambda \phi_0^2 + \frac{\lambda}{32\pi^2} \left\{ \left[ m_{ph}^2 + \left(\xi_R - \frac{1}{6} \right) R \right] \ln \left( \frac{m_{ph}^2}{\tilde{\mu}^2} \right) \right.\nonumber\\
 &+&\left. \left(\xi_R - \frac{1}{6} \right) R - 2 F(m_{ph}^2,\{R\}) \right\}. 
 \label{phys-mass-fin}
\end{eqnarray} Here, the explicit dependence on the renormalization scale $\tilde{\mu}$  should be compensated with an implicit $\tilde{\mu}$-dependence on 
the finite MS parameters $m^2(\tilde{\mu})$, $\xi(\tilde{\mu})$ and $\lambda(\tilde{\mu})$. Indeed, the   invariance of this equation under changes of $\tilde{\mu}$  becomes manifest when  
we express it in terms of the renormalized quantities $m_R^2$, $\xi_R$ and $\lambda_R$. The latter are related to the former ones  by
\begin{subequations}
\begin{align}
m_R^2 &= \frac{m^2+\frac{\lambda}{16\pi^2}\left[ R_0 \frac{dF_{dS}}{dR}\Big|_{m_R^2,R_0}-F_{dS}(m_R^2,R_0) \right]}{\left[ 1 - \frac{\lambda}{32 \pi^2} \ln \left( \frac{m_R^2}{\tilde{\mu}^2} \right) \right]}, \label{mR-mu-dSren} \\
\left( \xi_R - \frac{1}{6} \right) &= \frac{\left( \xi - \frac{1}{6} \right) - \frac{\lambda}{16\pi^2} \frac{dF_{dS}}{dR}\Big|_{m_R^2,R_0} }{ \left[1 - \frac{\lambda}{32\pi^2} - \frac{\lambda}{32\pi^2} \ln \left( \frac{m_R^2}{\tilde{\mu}^2} \right) \right] }, \label{xiR-mu-dSren} \\
 \lambda_R &= \frac{\lambda}{\left[ 1 - \frac{\lambda}{32\pi^2} - \frac{\lambda}{32\pi^2} \ln \left( \frac{m_{R}^2}{\tilde{\mu}^2} \right) - \frac{\lambda}{32\pi^2} \left( \frac{(\xi_R - \frac{1}{6}) R_0}{m_R^2} -2 \frac{dF_{dS}}{d m_{ph}^2}\Big|_{m_R^2,R_0} \right) \right]}. \label{lambdaR-mu-dSren}
\end{align}
\label{ren-mu-dSren}
\end{subequations} Two useful $\tilde{\mu}$-independent combinations follow immediately from these relations:
\begin{equation}
 \frac{m_B^2}{\lambda_{B2}} = \frac{m^2}{\lambda} = \frac{m_R^2}{\lambda_R^*} + \frac{\left(\xi_R - \frac{1}{6} \right) R_0}{32 \pi^2}
 \label{m-lambda-dS}
\end{equation}
and
\begin{eqnarray}
 \frac{\left( \xi_B - \frac{1}{6} \right)}{\lambda_B} &=& \frac{\left( \xi - \frac{1}{6} \right)}{\lambda} \\
 &=& \frac{\left( \xi_R - \frac{1}{6} \right)}{\lambda_R} + \frac{\left( \xi_R - \frac{1}{6} \right)}{32\pi^2} \left[ \left( \xi_R - \frac{1}{6} \right) \frac{R_0}{m_R^2} - 2 \frac{dF_{dS}}{d m_{ph}^2}\Big|_{m_R^2,R_0} \right] + \frac{1}{16\pi^2} \frac{dF_{dS}}{dR}\Big|_{m_R^2,R_0} \notag \\
 &\equiv& \frac{\left( \xi_R - \frac{1}{6} \right)}{\lambda_R} + J(R_0,m_R^2,\xi_R). \notag
 \label{xi-lambda-dS}
\end{eqnarray}
where $\lambda_R^*$ is defined by
\begin{equation}
 \frac{1}{\lambda_R^*}\equiv\frac{1}{\lambda_R} +\frac{1}{32\pi^2}.
\end{equation}
Using these parameters, the self-consistent equation for $m_{ph}^2$ can be written as
\begin{eqnarray}
  m_{ph}^2& = &m_R^2 + \frac{\lambda_R^*}{2} \phi_0^2 + \frac{\lambda_R^*}{32\pi^2} \Biggl\{ \left[ m_{ph}^2 + \left(\xi_R - \frac{1}{6} \right) R \right] \ln \left( \frac{m_{ph}^2}{m_R^2} \right) \notag \\
  &+ &\left( m_{ph}^2 - m_R^2 \right) \left[ 2 \frac{dF_{dS}}{d m_{ph}^2}\Big|_{m_R^2,R_0} - \frac{(\xi_R - \frac{1}{6}) R_0}{m_R^2} \right] \label{mph-eq-dSren} \\ 
  &+& 2\left[ F_{dS}(m_R^2,R_0) + \frac{dF_{dS}}{dR}\Big|_{m_R^2,R_0} \left( R - R_0 \right) - F(m_{ph}^2,R) \right] \Biggr\}. \notag
\end{eqnarray}

Finally, as will be needed for the renormalization of the energy-momentum tensor in next section, we write the results for the counterterms associated to the non-MS renormalized parameters defined in Eq. \eqref{type-of-counterterms}:
\begin{eqnarray} 
 \delta \tilde{m}^2 &\equiv& m_B^2 - m_R^2 = -  \frac{m_B^2}{32\pi^2} \frac{m_R^2}{\left( \frac{m_R^2}{\lambda_R^*}+\frac{\left(\xi_R - \frac{1}{6} \right) R_0}{32 \pi^2} \right)} \left[ \frac{2}{\epsilon} + \ln \left( \frac{m_{R}^2}{\tilde{\mu}^2} \right) - 2\frac{dF_{dS}}{d m_{ph}^2}\Big|_{m_R^2,R_0} \right],\\
 \delta \tilde{\xi} &\equiv& \xi_B - \xi_R = - \frac{\left( \xi_B - \frac{1}{6} \right)}{32\pi^2} \frac{\left\{ \left( \xi_R - \frac{1}{6} \right) \left[ \frac{2}{\epsilon} + 1 + \ln \left( \frac{m_{R}^2}{\tilde{\mu}^2} \right) \right] + 2 \frac{dF_{dS}}{dR}\Big|_{m_R^2,R_0} \right\}}{\left[ \frac{\left( \xi_R - \frac{1}{6} \right)}{\lambda_R} + J \right]} , \label{deltatildexi} \\
  \delta \tilde{\lambda} &\equiv& \lambda_{B2} - \lambda_R = -  \frac{\lambda_{B2} \lambda_R}{32\pi^2} \left[ \frac{2}{\epsilon} + 1 + \ln \left( \frac{m_{R}^2}{\tilde{\mu}^2} \right) + \frac{(\xi_R - \frac{1}{6}) R_0}{m_R^2} -2 \frac{dF_{dS}}{d m_{ph}^2}\Big|_{m_R^2,R_0} \right].
\end{eqnarray}
Note that the well known one-loop results can be recovered from these expressions,  making the replacements $m_B^2 \to m_R^2$, $\xi_B \to \xi_R$, $\lambda_{B2} \to \lambda_R$, and $R_0\to 0$ on the right-hand-sides.

\section{Renormalization of the semiclassical Einstein equations}\label{SEESect}

So far we have dealt with Eqs. \eqref{field-eq} and \eqref{gap-eq}, that give the dynamics of $\phi_0$ and $G$ for a given choice
of metric $g_{\mu\nu}$. However these equations do not take into account the  effect of the quantum field on the background geometry. In order to assess  
whether this backreaction is important or not, we must deal with the SEE, obtained from the stationarity condition given in Eq. (\ref{SEEdef}) with the gravitational action Eq. (\ref{Sg}) and 
the definition of the  vacuum expectation value of the energy-momentum tensor given in Eq. (\ref{tmunu-def}).
The resulting equations are 
\begin{equation}
 \kappa_B^{-1} G_{\mu \nu} + \Lambda_B \kappa_B^{-1} g_{\mu \nu} + \alpha_{1B} \, ^{(1)}H_{\mu\nu} + \alpha_{2B} \, ^{(2)}H_{\mu\nu} + \alpha_{3B} \, H_{\mu\nu} = \langle T_{\mu\nu} \rangle, \label{SEE1}
\end{equation} 
where  $\kappa_B=8\pi G_B$. An explicit expression for the tensors  $^{(1,2)}H_{\mu\nu}$ and $H_{\mu\nu}$  can be found for instance in  \cite{mazzi-paz2}.

The renormalization procedure then involves the calculation of $ \langle T_{\mu\nu} \rangle$ and the regularization of its divergences. The divergences can be of either one of two types, independent of the field $\phi_0$ and therefore only geometrical, or otherwise $\phi_0$-dependent either explicitly or implicitly through $m_{ph}^2(\phi_0)$. The SEE are renormalizable if, with the same choice of counterterms as for the field and gap equations, the non-geometrical divergences can be completely dealt with.
In order to absorb the geometrical divergences in the renormalization of the parameters of the gravitational part of the action, $\kappa_B^{-1}$, $\Lambda_B$ and $\alpha_{iB}$, these 
divergences must be proportional to the tensors that appear on the left-hand side of Eq. \eqref{SEE1} (note that in four spacetime dimensions the tensors  $^{(1,2)}H_{\mu\nu}$ and $H_{\mu\nu}$  are not all independent). 

We will follow the usual procedure and define the renormalized energy-momentum tensor as
\begin{equation}
\langle T_{\mu\nu} \rangle_{ren}=\langle T_{\mu\nu} \rangle - \langle T_{\mu\nu} \rangle_{ad4}\, ,
\end{equation}
where the fourth adiabatic order is understood as the expansion containing up to four derivatives of the metric and up to two derivatives of the mean field
\cite{mazzi-paz2}. Our goal in this section is to show that with the choice of the counterterms for the field and gap equations,  $\langle T_{\mu\nu} \rangle_{ad4}$
only contains geometric divergences, that can be absorbed into the bare gravitational constants.

The expectation value $\langle T_{\mu\nu} \rangle$ can be formally computed from  the definition Eq. \eqref{tmunu-def}. One can show that \cite{Ramsey}
\begin{equation}
\langle T_{\mu\nu}\rangle = T_{\mu\nu}(\phi_0)+\langle T_{\mu\nu}^f \rangle + \frac{\lambda_{B2}}{32}[G_1]^2 g_{\mu\nu}, \label{Tmunu0}
\end{equation} 
where the first term is the classical  energy-momentum tensor evaluated at $\phi_0$
\begin{eqnarray}
T_{\mu\nu} (\phi_0) = -\frac{2}{\sqrt{-g}}\frac{\delta S_m}{\delta g^{\mu\nu}} &=& (1-2\xi_B)\phi_{0,\mu}\phi_{0,\nu}-2\xi_B\phi_{0;\mu\nu}\phi_0+2\xi_B g_{\mu\nu}{\phi_0\square \phi_0}+\xi_B \phi_0^2G_{\mu\nu}\nonumber\\
&&+\left(2\xi_B-\frac{1}{2}\right) g_{\mu\nu}\phi_0^{,\lambda}\phi_{0,\lambda} -\frac{m_B^2}{2} g_{\mu\nu}\phi_0^2-\frac{\lambda_{B4}}{4!} g_{\mu\nu}\phi_0^4. \label{Tmunu-cl}
\end{eqnarray}
The second term  is formally the mean value of the energy-momentum tensor of a free field, constructed with the two-point function $G_1$. 
More explicitly, it can be written  as \cite{Synge,mazzi-paz2}
\begin{equation}
\langle T_{\mu\nu}^f \rangle = -\frac{1}{2}[{{G_1}_{;\mu\nu}}]+\frac{\left(1-2\xi_B\right)}{4} {{[G_1]}_{;\mu\nu}}+\left(\xi_B-\frac{1}{4}\right)\frac{g_{\mu\nu}}{2}{{\square[G_1]}} +\xi_B R_{\mu\nu}\frac{[G_1]}{2}. \label{Tmunu2} 
\end{equation}

As a side point, we mention that one could also derive Eq. (\ref{Tmunu0}) using a different approach:   take the classical  energy-momentum tensor for the action Eq. \eqref{classical-S},
evaluate for $\phi = \phi_0 + \varphi$ and then expand on the fluctuation $\varphi$. Afterwards take the expectation value $\langle \dots \rangle$ and recall that in the Hartree approximation  
one can write the expectation values of products of fields in terms of $\phi_0$ and $\langle \varphi^2 \rangle = [G_1]/2$ (and derivatives), using that
\begin{subequations}
\begin{align}
\langle \varphi^3 \rangle &= 0, \\
\langle \varphi^4 \rangle &= \frac{3}{4}[G_1]^2. 
\end{align}
\end{subequations}

For the renormalization it is useful to separate, in the expressions for $T_{\mu\nu}(\phi_0)$ and $\langle T_{\mu\nu}^f \rangle$,  the bare couplings into the corresponding  renormalized parts and the nonminimal subtraction counterterms
\begin{eqnarray}
T_{\mu\nu}(\phi_0)&=&T_{\mu\nu}(\phi_0)\Bigg|_{B=R}+ \delta\tilde{\xi} \left( - {\phi_0^2}_{;\mu\nu} + g_{\mu\nu} \square \phi_0^2 + \phi_0^2 G_{\mu\nu}  \right) - \frac{\delta \tilde{m}^2}{2} \phi_0^2 g_{\mu\nu} \label{TmunuA}\\
\langle T_{\mu\nu}^f \rangle&=&\langle T_{\mu\nu}^f \rangle \Bigg|_{B=R} + \frac{\delta\tilde{\xi}}{2} \left( - {{[G_1]}_{;\mu\nu}}+  g_{\mu\nu}{{\square[G_1]}} + R_{\mu\nu}[G_1] \right), \label{TmunuB2}
\end{eqnarray}
where $B=R$ is a notational shorthand to indicate a replacement of the bare couplings with the renormalized ones. 
It will be also useful to write separately the interaction term in the classical energy momentum tensor
\begin{equation}
T_{\mu\nu}(\phi_0)\Bigg|_{B=R}=T_{\mu\nu}(\phi_0)\Bigg|_{B=R,free}\hspace{-0.4cm}-\frac{\lambda_{B4}}{4!} \phi_0^4 g_{\mu\nu}\, .
\end{equation}
 Note that while  there are no divergences in $T_{\mu\nu}(\phi_0) |_{B=R,free}$,  the quantity $\langle T_{\mu\nu}^f \rangle |_{B=R}$ still has divergences that arise from the coincidence limit of $G_1$ and of its derivatives.
Recall Eq. (\ref{gap-eq}), which implies that in our case the two-point function is that of a field of mass $m_{ph}^2$ and curvature coupling $\xi_R$.

We are now ready to show that the counterterms already chosen to renormalize the mean field and gap equations also cancel the non-geometrical divergences in $\langle T_{\mu\nu}\rangle$. 
The third term of Eq. \eqref{Tmunu0} as well as the terms that were isolated in Eq. \eqref{TmunuB2} involve $[G_1]$ and its derivatives, and therefore they 
can be expressed in terms of $m_{ph}^2$ and the bare couplings by using that the physical mass is defined by the equality of  Eqs.  \eqref{unrenprop} and  \eqref{gap-eq}, which in a more convenient form reads
\begin{equation}
\frac{\lambda_{B2}}{4}[G_1]=m_{ph}^2-\tilde{\delta \xi} R-m_B^2-\frac{\lambda_{B2}}{2}\phi_0^2 \label{G1-mph}.
\end{equation}
With this replacement we have
\begin{eqnarray}
\langle T_{\mu\nu}\rangle &=& T_{\mu\nu}(\phi_0) \Bigg|_{B=R,free} + \langle T_{\mu\nu}^f \rangle \Bigg|_{B=R} + \frac{(3\lambda_{B2} - \lambda_{B4})}{4!} \phi_0^4 g_{\mu\nu} \nonumber\\
&&+ \frac{2\delta\tilde{\xi}}{\lambda_{B2}} \left[ - {m_{ph}^2}_{;\mu\nu} + g_{\mu\nu} \square m_{ph}^2 + G_{\mu\nu} m_{ph}^2 \right] + \frac{m_{ph}^4}{2\lambda_{B2}} g_{\mu\nu} - m_{ph}^2  \frac{m_B^2}{\lambda_{B}}  g_{\mu\nu}\nonumber \label{Tmunu3} \\
&& + \frac{\delta\tilde{\xi}^2}{\lambda_{B2}} \, ^{(1)} H_{\mu\nu} - 2\delta\tilde{\xi} \frac{m_B^2}{\lambda_{B2}} G_{\mu\nu} + \frac{m_B^2}{2} \frac{m_B^2}{\lambda_{B}} g_{\mu\nu} \nonumber \\
&&+ (m_R^2 - m_{ph}^2)  \frac{\phi_0^2}{2}  g_{\mu\nu} .\end{eqnarray} 
Here the term proportional to $\phi_0^4$ is already finite because of the relation Eq. \eqref{lambdas2-4} between the counterterms, and thus equal to $\lambda_R \phi_0^4 g_{\mu\nu}/12$. The fourth, fifth and sixth terms contain the non-geometrical divergences that will have to be cancelled by those from $\langle T_{\mu\nu}^f \rangle|_{B=R}$. The remaining terms  contain purely geometrical divergences.  

It is worth to emphasize  that the divergences in Eq. \eqref{Tmunu3} are  proportional to
simple poles in $\epsilon$. Indeed, from the definition of $\delta\tilde{\xi} = \xi_B - \xi_R$ and the relations \eqref{xi-lambda-dS} it is straightforward to see that
\begin{subequations}
\begin{align}
 \frac{\delta\tilde{\xi}}{\lambda_{B2}} &= \left( \frac{1}{\lambda_R} - \frac{1}{\lambda_{B2}} \right) \left(\xi_R - \frac{1}{6} \right) + J, \\
 \frac{\delta\tilde{\xi}^2}{\lambda_{B2}} &= \lambda_{B2} \left[ \frac{\left( \xi_R - \frac{1}{6} \right)}{\lambda_R} + J \right]^2 - 2 \left( \xi_R - \frac{1}{6} \right) \left[ \frac{\left( \xi_R - \frac{1}{6} \right)}{\lambda_R} + J \right] \notag \\ 
 &+ \frac{\left( \xi_R - \frac{1}{6} \right)^2}{\lambda_{B2}},
\end{align}
\end{subequations}
which are exact expressions. Note that $\lambda_{B2}^{-1}$ contains just a simple pole,
\begin{equation}
 \frac{1}{\lambda_{B2}} = \frac{1}{\lambda} + \frac{1}{16\pi^2 \epsilon}. \label{lambdaB-ala-1}\, 
\end{equation}


We now expand $\langle T_{\mu\nu}\rangle$ up to the fourth adiabatic order.  We will use the explicit expressions for the coincidence limit of $G_1$ and its derivatives that are given in Ref. \cite{mazzi-paz2}. The fourth adiabatic order expansion for $\langle \tilde{T}_{\mu\nu} \rangle \equiv \langle T_{\mu\nu}^f \rangle|_{B=R}$ is 
\begin{eqnarray}\nonumber
\langle \tilde{T}_{\mu\nu} \rangle_{ad4} &=& \frac{1}{16\pi^2} \left( \frac{m_{ph}^2}{\mu^2} \right)^{\epsilon/2} \Biggl[ \frac{1}{2} m_{ph}^4 \, g_{\mu \nu} \, \Gamma\left( -2 -\frac{\epsilon}{2} \right) + m_{ph}^2 \left\{ \frac{1}{2} [\Omega_1] g_{\mu\nu} + \left(\xi_R-\frac{1}{6} \right) R_{\mu \nu} \right\} \\ 
&\times& \Gamma\left( -1 -\frac{\epsilon}{2} \right) + \left\{ \frac{1}{2} [\Omega_2] g_{\mu\nu} + \left(\xi_R-\frac{1}{6} \right) R_{\mu\nu} [\Omega_1] - [\Omega_{1;\mu\nu}] \right. \nonumber\\ &+& 
\left. \left(\frac{1}{2} - \xi_R \right) [\Omega_1]_{;\mu\nu} + \left(\xi_R - \frac{1}{4} \right) g_{\mu\nu} \, \square[\Omega_1] \right\} \, \Gamma\left(-\frac{\epsilon}{2} \right) \Biggr], \label{Tmunutilde1}
\end{eqnarray} where the expressions for $[\Omega_1]$, $[\Omega_2]$ and $[\Omega_{1;\mu\nu}]$ can be found in  the Appendix A of  \cite{mazzi-paz2}. Notice 
however that here these contributions are expressed in terms of $\xi_R$ instead of $\xi_B$. Expanding for $\epsilon \to 0$, regrouping the geometric terms to form the appropriate tensors and separating the divergent part one arrives at
\begin{eqnarray}
 \nonumber \langle \tilde{T}_{\mu\nu} \rangle_{ad4} &=& \frac{1}{16\pi^2 \epsilon} \Biggl\{ - \frac{1}{2} m_{ph}^4 g_{\mu\nu} + 2 m_{ph}^2 \left( \xi_R - \frac{1}{6} \right) G_{\mu \nu} + \frac{1}{90} \left[ ^{(2)}H_{\mu\nu} - H_{\mu\nu} \right]  \notag \\
&-&  \nonumber \left( \xi_R - \frac{1}{6} \right)^2 \, ^{(1)}H_{\mu\nu}+ 2\left(\xi_R - \frac{1}{6} \right) \left( g_{\mu\nu} \square m_{ph}^2 - {m_{ph}^2}_{;\mu\nu} \right) \Biggr\} \label{Tmunutilde3}\\
&+& \frac{m_{ph}^4}{64\pi^2} g_{\mu\nu} \left[ \frac{1}{2} - \ln \left( \frac{m_{ph}^2}{\tilde{\mu}^2} \right) \right] + \frac{m_{ph}^2}{16\pi^2} \left( \xi_R - \frac{1}{6} \right) G_{\mu \nu} \ln \left( \frac{m_{ph}^2}{\tilde{\mu}^2} \right) \nonumber \\
&+& \frac{1}{32\pi^2} \left[ \frac{1}{90} \left( ^{(2)}H_{\mu\nu} - H_{\mu\nu} \right) - \left( \xi_R - \frac{1}{6} \right)^2 \, ^{(1)}H_{\mu\nu}\right. \nonumber \\ 
&+& \left. 2\left(\xi_R - \frac{1}{6} \right) \left( g_{\mu\nu} \square m_{ph}^2 - {m_{ph}^2}_{;\mu\nu} \right) \right] \left[ 1 + \ln \left( \frac{m_{ph}^2}{\tilde{\mu}^2} \right) \right].
\end{eqnarray}
Replacing Eq. \eqref{Tmunutilde3} into Eq. \eqref{Tmunu3} one can verify that the non-geometrical divergences
in  Eq. \eqref{Tmunu3} cancel out. This result  shows the renormalizability of the SEE within the consistent renormalization approach.

In order to complete the analysis, we write the full expression for the fourth adiabatic order, which we separate in its divergent and a convergent parts:
\begin{equation}
\langle T_{\mu\nu} \rangle_{ad4} = \langle T_{\mu\nu} \rangle_{ad4}^{div} + \langle T_{\mu\nu} \rangle_{ad4}^{con}\, ,
\label{TmunuAd4}
\end{equation}
with
\begin{eqnarray}
\langle T_{\mu\nu} \rangle_{ad4}^{div} &=& \frac{1}{90} \frac{1}{32\pi^2} \left[ \frac{2}{\epsilon} + 1 + \ln \left( \frac{m_{R}^2}{\tilde{\mu}^2} \right) \right] \left( ^{(2)}H_{\mu\nu} - H_{\mu\nu} \right) - 2 \delta \tilde{\xi} \left[ \frac{m_R^2}{\lambda_R^*} + \frac{\left( \xi_R - \frac{1}{6} \right) R_0}{32\pi^2} \right] G_{\mu\nu}  \nonumber \\
&+& \delta \tilde{\xi} \left[ \frac{\left( \xi_R - \frac{1}{6} \right)}{\lambda_R} + J \right] \, ^{(1)}H_{\mu\nu} +  \frac{\delta \tilde{m}}{2} \left[ \frac{m_R^2}{\lambda_R^*} + \frac{\left( \xi_R - \frac{1}{6} \right) R_0}{32\pi^2} \right] g_{\mu\nu}- \frac{m_R^4}{64 \pi^2} g_{\mu\nu}  \, , 
\label{TmunuAd4div} 
\end{eqnarray}
and
\begin{eqnarray}
\langle T_{\mu\nu} \rangle_{ad4}^{con} &=& 
 T_{\mu\nu}(\phi_0) \Bigg|_{B=R,free} + \frac{\lambda_R}{12} \phi_0^4 \, g_{\mu\nu} + \left( \frac{m_R^2}{2} - m_{ph}^2 \right) \left[ \frac{m_R^2}{\lambda_R^*} + \frac{\left( \xi_R - \frac{1}{6} \right) R_0}{32\pi^2} \right] g_{\mu\nu} \notag \\
&+& \frac{m^4_{ph}}{64\pi^2} \left[ \frac{32\pi^2}{\lambda_R^*} + \frac{1}{2} + \left( \xi_R - \frac{1}{6} \right) \frac{R_0}{m_R^2} - 2 \frac{dF_{dS}}{d m_{ph}^2}\Big|_{m_R^2,R_0} \right] g_{\mu\nu}  \nonumber\\
&+& \frac{1}{16\pi^2} \left[ 2 m_{ph}^2 G_{\mu\nu} - \left( \xi_R - \frac{1}{6} \right) \, ^{(1)} H_{\mu\nu} + 2 g_{\mu\nu} \square m_{ph}^2 - 2 {m_{ph}^2}_{;\mu\nu} \right] \frac{dF_{dS}}{dR}\Big|_{m_R^2,R_0} \notag \\
&+& \frac{1}{32\pi^2} \Biggl\{ -\frac{m^4_{ph}}{2} g_{\mu\nu} + 2 m_{ph}^2 \left(\xi_R - \frac{1}{6} \right) G_{\mu\nu} + \frac{1}{90} \left( ^{(2)}H_{\mu\nu} - H_{\mu\nu} \right) \nonumber \\ 
&& ~~~~~~~~~ - \left( \xi_R - \frac{1}{6} \right)^2  \, ^{(1)} H_{\mu\nu}+ 2\left(\xi_R - \frac{1}{6} \right) \left( g_{\mu\nu} \square m_{ph}^2 - {m_{ph}^2}_{;\mu\nu} \right) \Biggr\} \ln \left( \frac{m_{ph}^2}{m_R^2} \right) \notag \\ 
&-& \frac{m_{ph}^2}{16\pi^2} \left(\xi_R - \frac{1}{6} \right) G_{\mu\nu} + \left(m_R^2 - m_{ph}^2 \right) \frac{\phi_0^2}{2} g_{\mu\nu} + \frac{m_R^4}{64 \pi^2} g_{\mu\nu}\, .  \label{TmunuAd4con} 
\end{eqnarray}
As anticipated, the divergent part contains purely geometric divergences. The convergent part is field dependent, finite, and written 
in terms of the renormalized parameters (therefore independent of $\tilde{\mu}$). 
To ensure the correct one-loop limit of the cosmological constant counterterm, we included the finite contribution $- \frac{m_R^4}{64 \pi^2} g_{\mu\nu}$ in $\langle T_{\mu\nu} \rangle_{ad4}^{div}$.  

Now we can add and subtract $\langle T_{\mu\nu} \rangle_{ad4}$ in the right-hand side of the SEE 
\begin{eqnarray}
&& \kappa_B^{-1} (G_{\mu \nu} + \Lambda_B g_{\mu \nu}) + \alpha_{1B} \, ^{(1)}H_{\mu\nu} + \alpha_{2B} \, ^{(2)}H_{\mu\nu} + \alpha_{3B} \, H_{\mu\nu} = 
\nonumber\\ && \left[ \langle T_{\mu\nu} \rangle - \langle T_{\mu\nu} \rangle_{ad4} \right]
 + \langle T_{\mu\nu} \rangle_{ad4}^{div} + \langle T_{\mu\nu} \rangle_{ad4}^{con}\, ,
\end{eqnarray}
where the quantity between square brackets on the right-hand side is defined as $\langle T_{\mu\nu} \rangle_{ren}$. Renormalization is completed by absorbing $\langle T_{\mu\nu} \rangle_{ad4}^{div}$ into  a redefinition of the bare gravitational constants of the left-hand side. Then the renormalized gravitational parameters read
\begin{subequations}
\begin{align}
 \kappa_B^{-1} &= \kappa_R^{-1} + \frac{m_B^2}{8\pi^2} \left\{ \left(\xi_R - \frac{1}{6} \right) \left[ \frac{1}{\epsilon} + \frac{1}{2} + \frac{1}{2} \ln \left( \frac{m_{R}^2}{\tilde{\mu}^2} \right) \right] - \frac{dF_{dS}}{dR}\Big|_{m_R^2,R_0} \right\}, \\
 \Lambda_B \kappa_B^{-1} &= \Lambda_R \kappa_R^{-1} - \frac{m_B^2 m_R^2}{32\pi^2} \left[ \frac{1}{\epsilon} + \frac{1}{2} \ln \left( \frac{m_{R}^2}{\tilde{\mu}^2} \right) - \frac{dF_{dS}}{d m_{ph}^2}\Big|_{m_R^2,R_0} \right] - \frac{m_R^4}{64 \pi^2}, \\
 \alpha_{1B} &= \alpha_{1R} - \frac{\left( \xi_B - \frac{1}{6} \right)}{16\pi^2} \left\{ \left(\xi_R - \frac{1}{6} \right) \left[ \frac{1}{\epsilon} + \frac{1}{2} + \frac{1}{2} \ln \left( \frac{m_{R}^2}{\tilde{\mu}^2} \right) \right] - \frac{dF_{dS}}{dR}\Big|_{m_R^2,R_0} \right\}, \\
 \alpha_{2B} &= \alpha_{2R} + \frac{1}{1440 \pi^2} \left[ \frac{1}{\epsilon} + \frac{1}{2} + \frac{1}{2} \ln \left( \frac{m_{R}^2}{\tilde{\mu}^2} \right) \right], \\
 \alpha_{3B} &= \alpha_{3R} - \frac{1}{1440 \pi^2} \left[ \frac{1}{\epsilon} + \frac{1}{2} + \frac{1}{2} \ln \left( \frac{m_{R}^2}{\tilde{\mu}^2} \right) \right].
 \end{align}
\end{subequations}
These are consistent with the well known one-loop results when replacing the bare parameters in the right-hand side (in the counterterms) by the renormalized ones and setting $R_0 \to 0$, thus justifying the choice of $\langle T_{\mu\nu} \rangle_{ad4}^{div}$ in Eq. \eqref{TmunuAd4div}. As it happens for the field parameters, the relation between the bare and renormalized expressions is $\tilde{\mu}$-dependent. 

Finally, the renormalized SEE are
\begin{equation}
 \kappa_R^{-1} G_{\mu \nu} + \Lambda_R \kappa_R^{-1} g_{\mu \nu} + \alpha_{1R} \, ^{(1)}H_{\mu\nu} + \alpha_{2R} \, ^{(2)}H_{\mu\nu} + \alpha_{3R} \, H_{\mu\nu} = \langle T_{\mu\nu} \rangle_{ren} + \langle T_{\mu\nu} \rangle_{ad4}^{con}, \label{SEEren}
\end{equation} 
which, as expected, are expressed only in terms of renormalized parameters. 

\section{Interacting fields in de Sitter spacetime}\label{DSSect}

In this section we apply the previous results to de Sitter spacetime with  $ds^2 = - dt^2 + e^{2Ht} d\vec{x}^2$, and
compute explicitly the renormalized energy-momentum tensor, and the SEE.  We then
consider both the field equation and the SEE to analyze the existence of self-consistent solutions.

\subsection{Gap and semiclassical Einstein equations}

In de Sitter spacetime, the solution of the Eq.  (\ref{gap-eq}) for the propagator, which is the one of a free field with mass $m_{ph}^2$,  is known exactly for an arbitrary number of dimensions $n$.
The expression for the coincidence limit $[G_1]$ is
\begin{equation}
 [G_1] = \frac{2 H^{n-2}}{(4\pi \mu^2)^{n/2}} \, \Gamma \left( 1 - \frac{n}{2} \right) \frac{\Gamma \left( \frac{n-1}{2} + \nu_n \right) \, \Gamma \left( \frac{n-1}{2} - \nu_n \right)}{\Gamma \left( \frac{1}{2} + \nu_n \right) \, \Gamma \left( \frac{1}{2} - \nu_n \right)},
\end{equation}
where $\nu_n^2 = \frac{(n-1)^2}{4} - \frac{m_{ph}^2}{H^2} - \xi_R n(n-1)$ and $R=n(n-1) H^2$. 

To make use of the results of  previous sections we need to extract the function $F_{dS}(m_{ph}^2,R)$, defined in Eq. \eqref{adiab-G1}, from this exact expression. For this, we  set $n=4+\epsilon$ and expand for $\epsilon \to 0$, holding $R$ fixed.  Doing this, as shown in detail in paper I,  we obtain  the following  expression for the function $F(m_{ph}^2,\{R\})$ in de Sitter spacetime 
\begin{eqnarray}
F_{dS}(m_{ph}^2,R) = R \, f(m_{ph}^2/R) = -\frac{R}{2} &\Biggl\{& \left(\frac{m_{ph}^2}{R} + \xi_R -\frac{1}{6} \right) \left[ \ln\left( \frac{R}{12 m_{ph}^2} \right) + g\left( m_{ph}^2/R + \xi_R \right) \right] \notag \\
&&- \left( \xi_R -\frac{1}{6} \right) - \frac{1}{18} \Biggr\} \label{F-dS} 
\end{eqnarray}
with
\begin{equation}
g(y) \equiv \psi_{+} + \psi_{-} = \psi\left( \frac{3}{2} + \nu_4(y) \right) + \psi\left( \frac{3}{2} - \nu_4(y) \right),
\label{little-g} 
\end{equation}
and $R=12 H^2$, $\psi(x) = \Gamma^{'}(x)/\Gamma(x)$ is the digamma function and $\nu_4(y) = \sqrt{9/4-12y}$. From this equation one can check that  this function has all the expected properties: it is written only in terms of renormalized parameters, it is independent of $\epsilon$ and $\tilde{\mu}$, and it satisfies the correct limits Eqs. \eqref{F-prop1}, \eqref{F-prop2} and \eqref{F-prop3}. 

Therefore, the renormalized equation for the physical mass   $m_{ph}^2$  we are going to solve self-consistently together with the SEE we calculate below, can be written as:
\begin{eqnarray}
  m_{ph}^2 &=& m_R^2 + \frac{\lambda_R^*}{2} \phi_0^2 + \frac{\lambda_R^*}{32 \pi^2} \left[ m_{ph}^2 + \left( \xi_R - \frac{1}{6} \right) R \right] \left[ \ln \left( \frac{R}{12 m_{R}^2} \right) + g\left( m_{ph}^2/R + \xi_R \right) \right] \nonumber\\
  &-& \frac{\lambda_R^*}{32 \pi^2} \left( \xi_R - \frac{1}{9} \right) R . \label{mph-eq-dS}
\end{eqnarray}


In de Sitter spacetime all geometrical quantities  can be written in terms of only $R$ and $g_{\mu\nu}$. In $n$ dimensions they are:
\begin{subequations}
\begin{align}
R_{\mu \nu} &= \frac{R}{n} g_{\mu \nu}, \\
G_{\mu \nu} &= \left( \frac{1}{n} - \frac{1}{2} \right) R \, g_{\mu \nu}, \\
^{(1)} H_{\mu \nu} &=  \frac{1}{2} \left( 1 - \frac{4}{n} \right) R^2 \, g_{\mu \nu}, \\
^{(2)} H_{\mu \nu} &=  \frac{1}{2n} \left( 1 - \frac{4}{n} \right) R^2 \, g_{\mu \nu}, \\
    H_{\mu \nu} &=  \frac{1}{n(n-1)} \left( 1 - \frac{4}{n} \right) R^2 \, g_{\mu \nu}.
\end{align}
\label{deSitterGeom}
\end{subequations}
In fact, any 2nd-rank tensor is proportional to the metric, so that
\begin{equation}
 [{G_1}_{;\mu \nu}] = \frac{1}{n} [\square G_1] \, g_{\mu \nu}.
\end{equation}
De Sitter invariance also implies that any scalar function has vanishing derivative, and in particular that $[G_1]$ is independent of spacetime coordinates. The energy-momentum tensor will also be proportional to $g_{\mu \nu}$. Indeed, from the general expression Eq. \eqref{Tmunu0} together with Eqs. \eqref{Tmunu-cl} and \eqref{Tmunu2},  and using  Eq. \eqref{deSitterGeom}, we obtain
\begin{eqnarray}
\langle T_{\mu\nu}\rangle &=& \Biggl[-\frac{m^2_B}{2} \phi_0^2-\frac{\lambda_{B4}}{4!} \phi_0^4+\xi_B\phi_0^2 \left( \frac{1}{n} - \frac{1}{2} \right) R  -\frac{1}{2n}[\square G_1] - \frac{m^2_B}{4} [G_1]  \nonumber\\ 
&&+ \frac{1}{4} [{\square G_1}]+\xi_B \frac{[G_1]}{2} \left( \frac{1}{n} - \frac{1}{2} \right) R  - \frac{\lambda_{B2}}{8}\phi_0^2 [G_1]-\frac{\lambda_{B2}}{32}[G_1]^2 \Biggr] \, g_{\mu \nu}. 
\end{eqnarray} 
Once again we use Eq. \eqref{G1-mph} to make previous expression simpler, and we put $n=4+\epsilon$,
\begin{eqnarray}\nonumber
\langle T_{\mu\nu}\rangle &=& \Biggl\{-\frac{m^2_B}{2} \phi_0^2 - \frac{\xi_B}{4}\phi_0^2 R - \frac{\lambda_{B4}}{4!} \phi_0^4 - \frac{1}{8} \left[ m_B^2 + \frac{\lambda_{B2}}{2} \phi_0^2 \right] [G_1] \\
 &+& \frac{1}{4} \left( \frac{4}{4+\epsilon} -1 \right) \left[ \xi_B\phi_0^2 R -\frac{1}{2} \left( m_{ph}^2 - \delta \tilde{\xi} R \right) [G_1] \right] \Biggr\} \, g_{\mu \nu}.
\end{eqnarray}

Here we cannot set $\epsilon=0$ in the denominator yet, as it is multiplied by both the bare parameters and $[G_1]$, that contain poles in $\epsilon$ that could give finite terms. 
After some manipulations and dropping terms that vanish for $\epsilon \to 0$, it reads
\begin{eqnarray}
\langle T_{\mu\nu}\rangle &=& \Biggl\{ \frac{1}{2} \left[ \delta \tilde{m}^2 + \left( 1 +  \frac{\epsilon}{4 + \epsilon} \right) \delta \tilde{\xi} R \right] \left( \frac{m_R^2}{\lambda_R^*} + \frac{\left(\xi_R - \frac{1}{6} \right) R_0}{32 \pi^2} \right)\\
& +& \left( \frac{4}{4 + \epsilon} \right) \frac{\epsilon \, \delta \tilde{\xi}}{8} \left(\frac{\left( \xi_R - \frac{1}{6} \right)}{\lambda_R} + J \right) R^2 + \frac{1}{2} \left( \frac{m_R^2}{\lambda_R^*} + \frac{\left(\xi_R - \frac{1}{6} \right) R_0}{32 \pi^2} \right) \left( m_R^2 - m_{ph}^2 \right)\notag \\
&-& \frac{1}{4} (m_{ph}^2 + \xi_R R) \phi_0^2 + \frac{\lambda_R}{12} \phi_0^4+ \frac{1}{128 \pi^2} \left[ m_{ph}^2 + \left( \xi_R - \frac{1}{6} \right) R \right]^2  \Biggr\} g_{\mu \nu}.\notag
\end{eqnarray}

To compute the renormalized expectation value,  $ \langle T_{\mu \nu} \rangle_{ren} = \langle T_{\mu \nu} \rangle - \langle T_{\mu \nu} \rangle_{ad4}$,  we evaluate $\langle T_{\mu \nu} \rangle_{ad4}$
(given in Eq. \eqref{TmunuAd4}) in de Sitter spacetime,  using the $n$-dimensional geometrical expressions Eq. \eqref{deSitterGeom}. 
Separating the result again in $\langle T_{\mu\nu} \rangle_{ad4} = \langle T_{\mu\nu} \rangle^{div}_{ad4}  + \langle T_{\mu\nu} \rangle^{con}_{ad4} $,  up to order $\epsilon$ these two terms read 
\begin{eqnarray}\nonumber
 \langle T_{\mu\nu} \rangle_{ad4} ^{div} &=& \Biggl\{ \frac{1}{64\pi^2} \frac{R^2}{2160} + \frac{1}{2} \left[ \delta \tilde{m}^2 + \left( 1 +  \frac{\epsilon}{4 + \epsilon} \right) \delta \tilde{\xi} R \right] \left( \frac{m_R^2}{\lambda_R^*} + \frac{\left(\xi_R - \frac{1}{6} \right) R_0}{32 \pi^2} \right) \\
 &&+ \left( \frac{4}{4 + \epsilon} \right) \frac{\epsilon \, \delta \tilde{\xi}}{8} \left( \frac{\left( \xi_R - \frac{1}{6} \right)}{\lambda_R} + J \right) R^2 - \frac{m_R^4}{64 \pi^2} \Biggr\} \, g_{\mu \nu}, \label{TmunuAd4DS-div} \\
\nonumber \langle T_{\mu\nu} \rangle_{ad4} ^{con} &=& \Biggl\{ \frac{m_R^2}{2} \left[ \frac{m_R^2}{\lambda_R^*} + \frac{\left(\xi_R - \frac{1}{6} \right) R_0}{32 \pi^2} + \frac{m_R^2}{32 \pi^2} \right] + \frac{m_{ph}^2}{64 \pi^2} \left( \xi_R - \frac{1}{6} \right) R \\
&+& \frac{m_{ph}^4}{64\pi^2} \left[ \frac{32\pi^2}{\lambda_R^*} + \frac{1}{2} + \frac{(\xi_R - \frac{1}{6}) R_0}{m_R^2} -2 \frac{dF_{dS}}{d m_{ph}^2}\Big|_{m_R^2,R_0} \right] \notag \\
&-& \frac{m_{ph}^2}{64\pi^2} \left[ m_{ph}^2 + \left( \xi_R - \frac{1}{6} \right) R \right] \ln \left( \frac{m_{ph}^2}{m_R^2} \right) - \frac{m_{ph}^2 R}{32\pi^2} \frac{dF_{dS}}{d R}\Big|_{m_R^2,R_0} \notag \\
 &-& m_{ph}^2  \left[ \frac{m_R^2}{\lambda_R^*} + \frac{\left(\xi_R - \frac{1}{6} \right) R_0}{32 \pi^2} \right] - \left( m_{ph}^2 + \frac{\xi_R}{2} R \right) \frac{\phi_0^2}{2} + \frac{\lambda_R}{12} \phi_0^4 \Biggr\} \, g_{\mu \nu} .\label{TmunuAd4DS-con}
\end{eqnarray}
The first term of Eq. \eqref{TmunuAd4DS-div} is finite and is the source of the trace anomaly \cite{birrell}. Then we have 
\begin{eqnarray}
\langle T_{\mu \nu} \rangle_{ren} &=& - \frac{1}{64\pi^2} \Biggl\{ m_{ph}^2  \Biggl[ \left( \frac{32\pi^2}{\lambda_R^*} + \frac{(\xi_R - \frac{1}{6}) R_0}{m_R^2} -2 \frac{dF_{dS}}{d m_{ph}^2}\Big|_{m_R^2,R_0} \right)  (m_{ph}^2 - m_R^2)  - 16 \pi^2 \phi_0^2 \nonumber \\
&-& \left[ m_{ph}^2 + \left( \xi_R - \frac{1}{6} \right) R \right] \ln \left( \frac{m_{ph}^2}{m_R^2} \right) - 2 R  \frac{dF_{dS}}{d R}\Big|_{m_R^2,R_0} - 2 m_R^2 \frac{dF_{dS}}{d m_{ph}^2}\Big|_{m_R^2,R_0} \Biggr] \nonumber \\
&-& \frac{1}{2} \left( \xi_R - \frac{1}{6} \right)^2 R^2 + \frac{R^2}{2160} \Biggr\} g_{\mu \nu}. \label{TmunuRenDS}
\end{eqnarray}
To make contact with the known free and one-loop expressions, we use Eq.  \eqref{mph-eq-dSren} to arrive at a more familiar result
\begin{eqnarray}
\langle T_{\mu \nu} \rangle_{ren} &=& - \frac{1}{64\pi^2} \Biggl\{ 2 m_{ph}^2 \left[ F_{dS}(m_R^2,R_0) - R_0 \frac{dF_{dS}}{dR}\Big|_{m_R^2,R_0} - m_R^2 \frac{dF_{dS}}{dm_{ph}^2}\Big|_{m_R^2,R_0} - F(m_{ph}^2,R) \right] \nonumber \\
&-& \frac{1}{2} \left( \xi_R - \frac{1}{6} \right)^2 R^2 + \frac{R^2}{2160} \Biggr\} g_{\mu \nu} .\label{TmunuRenDS2}
\end{eqnarray}
Setting $R_0 \to 0$ and using Eq. \eqref{F-dS} for $F_{dS}$, gives an  expression that is exactly the same as in the one-loop calculation \cite{mazzi-paz2}, provided $m_{ph}^2 = m_R^2 + \lambda_R \phi_0^2/2$ instead of being the solution of the self-consistent Eq. \eqref{mph-eq-dS}. Furthermore, it is straightforward that the usual free field limit \cite{birrell} is satisfied, as $m_{ph}^2 \to m_R^2$ when $\lambda_R \to 0$. 
 
Turning finally to the SEE,  on the right-hand side we have 
\begin{eqnarray}
\langle T_{\mu \nu} \rangle_{ren} + \langle T_{\mu \nu} \rangle_{con}&=& \Biggl[ - \frac{1}{4} (m_{ph}^2 + \xi_R R) \phi_0^2 + \frac{\lambda_R}{12} \phi_0^4 \Biggr] g_{\mu \nu} \\
&-& \frac{1}{64\pi^2} \left\{ 32\pi^2 \left( \frac{m_R^2}{\lambda_R^*} + \frac{\left(\xi_R - \frac{1}{6} \right) R_0}{32 \pi^2} \right) \left( m_{ph}^2 - m_R^2 \right) \right.  \notag\\
&-&\left. m_R^4 + \frac{R^2}{2160} - \frac{1}{2} \left[ m_{ph}^2 + \left( \xi_R - \frac{1}{6} \right) R \right]^2 \right\} g_{\mu \nu},  \notag
\end{eqnarray}  while on the left-hand side we have $G_{\mu\nu} + \Lambda_R g_{\mu\nu} = (-R/4+\Lambda_R) g_{\mu\nu}$, as the quadratic tensors $^{(1)}H_{\mu\nu}$, $^{(2)}H_{\mu\nu}$ and $H_{\mu\nu}$ vanish for $n=4$. Then, canceling the $g_{\mu\nu}$ that appears on both sides, we have:
\begin{eqnarray}
M_{pl}^2 \left( -\frac{R}{4} + \Lambda_R \right)& =& - \frac{1}{8\pi} \left\{ \frac{R^2}{2160} + 32\pi^2 \left( \frac{m_R^2}{\lambda_R^*} + \frac{\left(\xi_R - \frac{1}{6} \right) R_0}{32 \pi^2} \right) \left( m_{ph}^2 - m_R^2 \right) - m_R^4  \right.\label{SEE} \\
&+& \left. 16 \pi^2 (m_{ph}^2 + \xi_R R) \phi_0^2 - 64 \pi^2 \frac{\lambda_R}{12} \phi_0^4 - \frac{1}{2} \left[ m_{ph}^2 + \left( \xi_R - \frac{1}{6} \right) R \right]^2  \right\}, \notag
\end{eqnarray}
where $M_{pl}$ is Planck's mass, and $\kappa_R = 8\pi/M_{pl}^2$.

\subsection{Self-consistent de Sitter solutions}

The back-reaction problem consists in solving simultaneously the mean field Eq. \eqref{field-eq}, the $m_{ph}^2$ Eq. \eqref{mph-eq-dS} and the SEE \eqref{SEE} self-consistently for the mean field $\phi_0$, the physical mass $m_{ph}^2$ and the scalar curvature of de Sitter spacetime $R$. This is a closed system of equations for a given set of parameters $m_R^2$, $\xi_R$, $\lambda_R$ and $\Lambda_R$, whose physically interesting solutions in a cosmological scenario are those with both $R$ and $\mathcal{M}_{ph}^2 = m_{ph}^2 + \xi_R R$ positive. The second condition comes from the fact that $\mathcal{M}_{ph}^2$ is the mass of the propagator, and it is a well known fact that the equation 
\begin{equation}
 \square G_1(x,x') = 0,
\end{equation}
has no de Sitter invariant solutions.

The gap Eq. \eqref{mph-eq-dS} is in itself a self-consistent equation for $m_{ph}^2(\phi_0,R)$, at fixed $\phi_0$ and $R$. Following paper I, in the small mass approximation ($y \equiv \mathcal{M}_{ph}^2/R \ll 1$) we have $g(y) \simeq - 1/4y + 11/6 - 2 \gamma_E + 49y/9$ an thus the gap equation becomes a quadratic equation for $y$,
\begin{equation}
 A_{dS} \, y^2 + \left[ B_{dS} - \frac{\lambda_R \phi_0^2}{2R} \right]\, y + C_{dS} = 0
 \label{gap-quadratic}
\end{equation}
where the coefficients are
\begin{subequations}
\begin{align}
A_{dS} =&~ 1 - \frac{\lambda_R}{32\pi^2} \left[ a\left( \frac{R}{R_0} \right) - g(y_0) - \left(y_0 - \frac{1}{6} \right) g'(y_0) - \frac{49}{54} \right], \\
B_{dS} =& - \left[ \frac{R_0}{R} y_0 + \xi_R \left( 1 - \frac{R_0}{R} \right) \right]  \nonumber\\
 &+ \frac{\lambda_R}{32\pi^2} \left\{ \frac{1}{4} +  \frac{1}{6} \left[ a\left( \frac{R}{R_0} \right) - g(y_0) - \left(y_0 - \frac{1}{6} \right) g'(y_0) \right] \right.\nonumber \\
 &~~~~~~~~~~~~ + \left. \left( 1 - \frac{R_0}{R} \right) \left(y_0 - \frac{1}{6} \right) - \left(y_0 - \frac{1}{6} \right)^2 g'(y_0) \right\}, \\
C_{dS} =& -\frac{\lambda_R}{768\pi^2}, 
\end{align}
\label{CuadEq-dSren}
\end{subequations}
with
\begin{equation}
a(x) \equiv 11/6 - 2\gamma_E + \ln(x),
\label{a-definition}
\end{equation}
and $y_0 = y(\phi_0=0,R=R_0) = m_R^2/R_0 + \xi_R$. The solution can be expressed analytically 
\begin{equation}
\mathcal{M}_{ph}^2 (\phi_0,R) = \frac{ -( R\, B_{dS} - \frac{\lambda_R \phi_0^2}{2}) + \sqrt{ \left[ R B_{dS} - \frac{\lambda_R \phi_0^2}{2} \right]^2 - 4 R^2\, A_{dS} C_{dS}  }  }{2 A_{dS}}. 
\label{Mph-phi-R}
\end{equation}
Here the ``plus'' branch was selected as the only real and positive solution (under the assumption that both $A_{dS} > 0$ and $B_{dS} > 0$, see paper I). This solution shall then be inserted into the mean field Eq. \eqref{field-eq}, which in de Sitter spacetime reads
\begin{equation}
\frac{dV_{eff}}{d\phi_0} \Bigg|_{\bar{\phi}_0} = \left( \mathcal{M}_{ph}^2(\bar{\phi_0},R) - \frac{1}{3} \lambda_R \bar{\phi}_0^2 \right) \bar{\phi}_0 = 0.
\label{dS-field-eq}
\end{equation}
This equation admits both symmetric solutions with $\bar{\phi}_0 = 0$ and solutions that spontaneously break the $Z_2$ symmetry,
\begin{equation}
 \bar{\phi}_0^2 = \frac{3}{\lambda_R} \mathcal{M}_{ph}^2(\bar{\phi_0},R).
\label{Sol-Field-eq}
\end{equation}
In other words, the effective potential $V_{eff}(\phi_0,R)$ may have other extrema besides the one in $\phi_0=0$.  The analysis of the effective potential has been done in 
paper I.

Studying the full backreaction problem by including the SEE \eqref{SEE} brings a new parameter into play, namely the cosmological constant $\Lambda_R$, as well as a new mass scale $M_{pl}^2$. In paper I, $R$ was considered fixed (i.e. as a parameter) and the effective potential and its minima were studied in order to find values of the remaining parameters $m_R^2$, $\xi_R$, $\lambda_R$ and $R_0$ at which both symmetric and broken phase solutions exist. In the small mass approximation, this amounts to analyzing constrains on combinations of the coefficients $A_{dS}$, $B_{dS}$ and $C_{dS}$ as functions of the parameters. Considering $R$ to be fixed makes sense under the assumption that the effect of the quantum field on the background curvature is small, and therefore it is possible to decouple the SEE from the field and gap equations. If this is indeed the case, the value of $R$ becomes effectively  independent of $\phi_0$ and $m_{ph}^2$, and is simply given by the parameter $\Lambda_R$. 

The aim of this section is to find some examples of self-consistent solutions involving all three equations and all three degrees of freedom. To this end, we take as starting point some sets of values of the parameters $m_R^2$, $\xi_R$, $\lambda_R$ and $R_0$ that were already shown in paper I to allow both symmetric and broken phase solutions. Then we look for solutions of $\phi_0$, $m_{ph}^2$ and $R$ for various values of $\Lambda_R$ and analyze how these differ from the classical solution. If this difference is small, then the backreaction can be indeed ignored, otherwise it should be taken into account.

One further point of discussion is whether the parameters $R_0$ and $\Lambda_R$ should be related or not. If this were to be the case, a sensible way of fixing one given the other would be to use the classical solution $R_0 = 4 \Lambda_R$.

\subsubsection{Symmetric Phase}


As mentioned above, the effective potential always has an extreme in $\phi_0 = 0$. Furthermore, it is easily shown that it must be a minimum as a consequence of both the restriction given by the Hartree approximation that $\mathcal{M}_{ph}^2 > 0$, and the fact that 
\begin{equation}\label{mini}
\frac{d^2 V_{eff}}{d \phi_0^2} \Bigg|_{\bar{\phi}_0=0} = \mathcal{M}_{ph}^2(\bar{\phi}_0 = 0, R) > 0.
\end{equation}

We solve the system of equations by setting $\phi_0=0$ in Eq.  \eqref{Mph-phi-R} to obtain $\mathcal{M}_{ph}^2$ as a function only of $R$ and then substituing into the SEE \eqref{SEE} to obtain an equation of the form
\begin{equation}
 \Lambda_R = I_s(R).
\end{equation}
where $I_s$ depends also on the parameters $m_R^2$, $\xi_R$, $\lambda_R$ and $R_0$. The subindex {\it s} stands for symmetric.

\subsubsection{Broken Phase}

In this phase, the solution given in Eq.  \eqref{Sol-Field-eq} to the field equation already implies $\mathcal{M}_{ph}^2 > 0$. It is important to note that the reason why the $\bar{\phi}_0 \neq 0$ solutions are allowed is the presence of  the $\lambda_R$ term in Eq. \eqref{dS-field-eq}, which comes as a consequence of imposing the 2PI consistency relations. Otherwise, the absence of such term would require that for $\bar{\phi}_0 \neq 0$ we had $\mathcal{M}_{ph}^2 = 0$, and as mentioned before for that case  there is no  de Sitter invariant vacuum \cite{mazzi-paz}.

Replacing the non vanishing solution to the field Eq. \eqref{Sol-Field-eq} into the gap equation in its quadratic form Eq. \eqref{gap-quadratic} (small mass approximation), we obtain a new quadratic equation for the non symmetric extrema of the potential  $\bar{\phi_0}^2(R)$, namely
\begin{eqnarray}
\frac{\lambda_R}{3} \, \bar{\phi_0}^2(R) = \frac{ -( R\, B_{dS} - \frac{\lambda_R \bar{\phi_0}^2}{2}) \pm \sqrt{ \left[ R B_{dS} - \frac{\lambda_R \bar{\phi_0}^2}{2} \right]^2 - 4 R^2\, A_{dS} C_{dS}  }  }{2 A_{dS}}. 
\label{mph-phi-R-broken}
\end{eqnarray}
Both branches give a solution with $\mathcal{M}_{ph}^2 > 0$, the smaller being the maximum and the larger the minimum of the effective potential. 
Following the analysis described in paper I,  one can show that the condition for the existence of symmetry breaking solutions is 
\begin{equation}
 B_{dS} - 2 \sqrt{ \left( \frac{3}{2} - A_{dS} \right) |C_{dS}| } > 0.
 \label{symm-breaking-cond}
\end{equation}

Once again, replacing $\phi_0(R)$ and $\mathcal{M}_{ph}^2(R)$ into the SEE gives an equation of the form
\begin{equation}
 \Lambda_R = I_b(R).
\end{equation}
The subindex {\it b} stands for broken. Note that in general $I_b(R)$ will be different from $I_s(R)$.

\subsubsection{Results}

In what follows we present the results in terms of  the relative deviation $(R-R_{cl})/R_{cl}$ of the backreaction solutions $R$  with respect to the classical solution $R_{cl} = 4 \Lambda_R$ as a function of $\Lambda_R$, for both  the symmetric and broken phases, when they exist.

\begin{figure}[ht]
  \centering
  \includegraphics[width=1.0\textwidth]{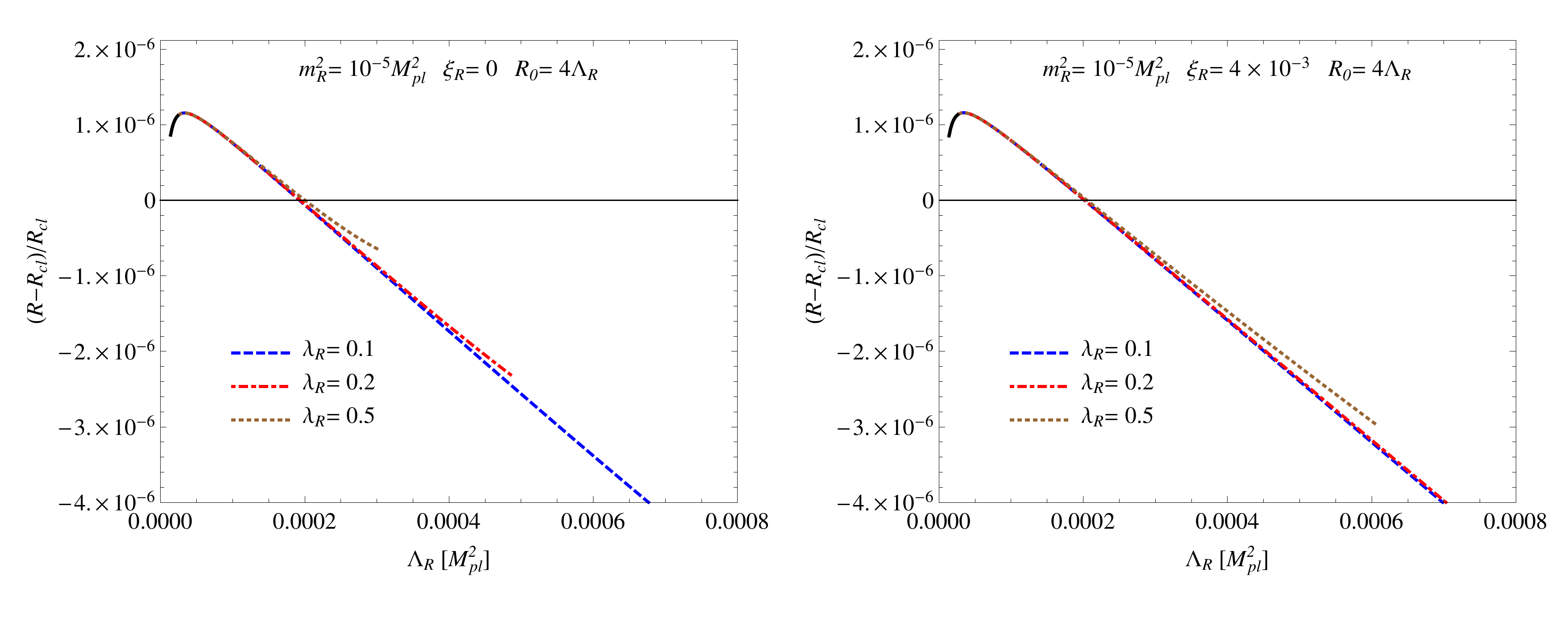}
  \caption{\small Relative deviation of the backreaction solution for the de Sitter spacetime curvature with respect to the classical solution, $(R-R_{cl})/R_{cl}$ as a function of  $\Lambda_R$ for different values of the coupling constant  $\lambda_R$.  The fixed parameters are $R_0=4\Lambda_R$,  $m_R^2 =  10^{-5} M_{pl}^2$. The left panel corresponds to  $\xi_R = 0$ and the right panel to $\xi_R=4\times 10^{-3}$. All curves correspond to the  symmetric phase (which is the only possible phase when $R_0=4\Lambda_R$). From bottom up: $\lambda_R=0.1$ (blue dashed line), $\lambda_R=0.2$ (red dotted-dashed line), $\lambda_R=0.5$ (brown dotted line). Notice that for small enough values of $\Lambda_R$ the curves are continued by black solid lines, indicating the regions where $\mathcal{M}_{ph}^2\geq R/10$.  }
  \label{fig:BR1}
\end{figure} 

 Let us first analyze the case where $R_0=4\Lambda_R$. This means that the renomalized parameters are defined at the value of scalar curvature of the background de Sitter spacetime that the theory would have had in the absence of backreaction. It is remarkable that in  this case no broken phase solutions exist. As an example, in Fig.  \ref{fig:BR1} we  have plotted the relative deviation for different values of the coupling constant $\lambda_R$, from bottom up: $\lambda_R=0.1,\,0.2$ and $0.5$,
with all curves corresponding to the symmetric phase and  $m_R^2=10^{-5}M_{pl}^2$.  On the left panel  the coupling to the curvature is minimal  $\xi_R=0$, while on the  right panel  $\xi_R=4\times 10^{-3}$.
 It is interesting to see that, due to the quantum corrections,  the curvature $R$ can be both larger or smaller than the classical one depending on the value of $\Lambda_R$. Notice that  solutions do not exist for all values of $\Lambda_R$.
 On the one hand, it can be seen that the approximation $\mathcal{M}_{ph}^2\ll R$ breaks down for small enough values of $\Lambda_R$. In order to make this explicit, in Fig.  \ref{fig:BR1}  and in the following,    black solid lines are used whenever $\mathcal{M}_{ph}^2\geq R/10$. On the other hand,  
 since we are considering only cases where the effective potential for $\phi_0$ is well defined, there is a ($\lambda_R$-dependent) lower bound for  the sum  $m_R^2/R+\xi_R$
 \cite{lower bound}, which will be violated for large enough values of $\Lambda_R$. 

\begin{figure}[ht]
  \centering
  \includegraphics[width=1.0\textwidth]{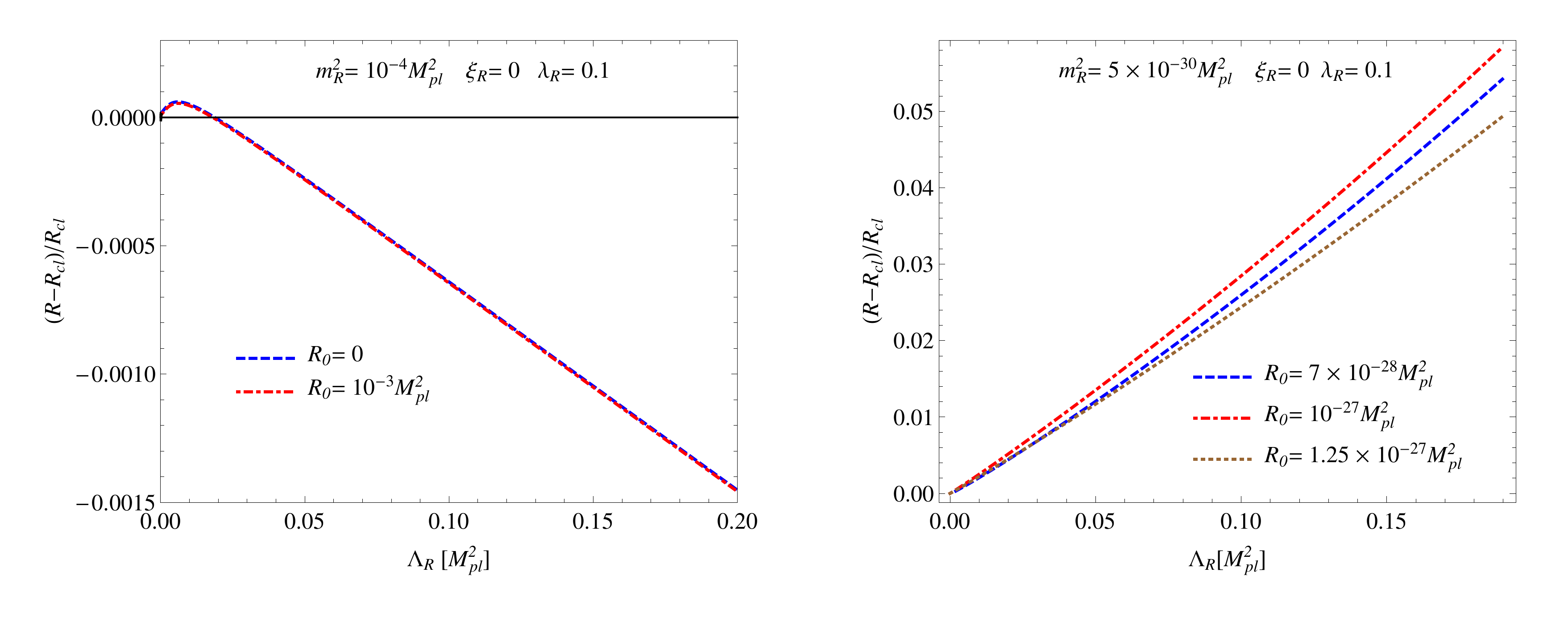}
  \caption{\small The same as in  figure \ref{fig:BR1}, but for different values of the curvature  $R_0$ associated to the renormalization point. Left panel:  symmetric phase solutions for $R_0= 0$ (blue dashed line) and $R_0=10^{-3} M_{pl}^2$ (red dotted-dashed line) with fixed parameters $m_R^2 =  10^{-4} M_{pl}^2$, $\xi_R = 0$, and $\lambda_R = 0.1$.  The curves are practically indistinguishable, illustrating  the solutions are quite independent on $R_0$.  Right panel: broken symmetry solutions  for $R_0=7\times 10^{-28} M_{pl}^2$  (blue dashed line),  $R_0=10^{-27} M_{pl}^2$  (red dotted-dashed line), and  $R_0=1.25\times 10^{-27} M_{pl}^2$  (brown dotted line) where the fixed parameters are $m_R^2 = 5 \times 10^{-30} M_{pl}^2$, $\xi_R = 0$, and $\lambda_R = 0.1$. In this case, the values of $R_0$ where  chosen to be in the small range 
  where a broken phase solution exists.}
  \label{fig:BR2}
\end{figure}   
 
Let us  now analyze  cases where $R_0$ is considered to be fixed and  independent of $\Lambda_R$.  In Fig.  \ref{fig:BR2},  the left panel  corresponds to the symmetric phase, while the right panel to the broken  one.  It can be seen that the backreaction is more significant in the broken  phase (e.g. the deviation is about 1\% for $\Lambda_R\simeq 0.04\, M_{pl}^2,\,R_0\simeq 10^{-27}M_{pl}^2$ and $m_{R}^2= 5\times 10^{-30}\, M_{pl}^2$),  while  in the symmetric phase the solution stays  closer to the classical one.
The difference between the backreaction and classical solutions may become important for larger values of the cosmological constant (not shown in the figure). Indeed, it can be shown that  the backreaction solution for the curvature $R$ vanishes  in the large (superplanckian) $\Lambda_R$ limit.  However,  adopting an effective field theory perspective,   
here  we are restricting the parameter space  to subplanckian values.   


As in general the broken phase solution is possible only for a suitable choice of the parameters \cite{Nos1},  in the right panel, the values of  $R_0$ had to be carefully chosen to be in the narrow window where  broken phase solutions exist, and they disappear below a small parameter-dependent value of $\Lambda_R$ (under $10^{-3} M_{pl}^2$ in the shown examples). One can verify that, depending on the values of the 
parameters, the approximation $\mathcal{M}_{ph}^2\ll R$ may break down. For the values considered in left panel of Fig.  \ref{fig:BR2} this happens for small enough values of $\Lambda_R$, while for the ones in the right panel the approximation remains valid.

 \newpage
 \begin{figure}[ht]
  \centering
  \includegraphics[width=1.0\textwidth]{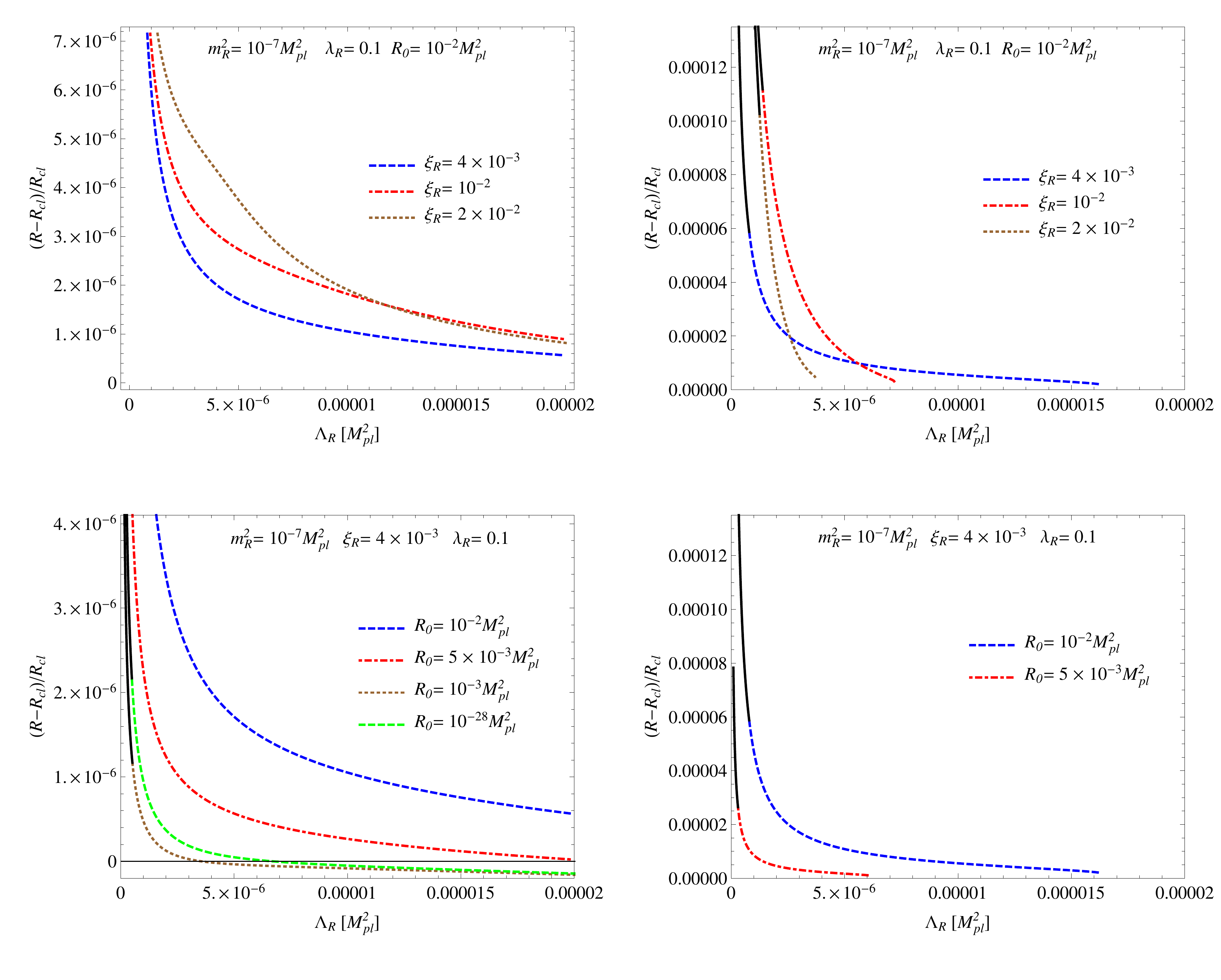}
  \caption{\small The deviation $(R-R_{cl})/R_{cl}$ vs. $\Lambda_R$ for the  backreaction solutions corresponding to the  symmetric  (on the left) and broken  (to the right) phases. Upper panels:  three curves corresponding to different values of the coupling to the curvature: $\xi_R= 4\times 10^{-3}$ (blue dashed line), $\xi_R= 10^{-2}$ (red dotted-dashed line), and $\xi_R= 2\times 10^{-2}$ (brown dotted line), where the fixed parameters are $m_R^2 = 10^{-7} M_{pl}^2$,  $\lambda_R=0.1$ and $R_0=10^{-2}M_{pl}^2$.  Lower panels: four different curves illustrating the dependence on the value of $R_0$ for $m_R^2 = 10^{-7} M_{pl}^2$, $\xi_R=4\times 10^{-3}$, and $\lambda_R=0.1$:  $R_0= 10^{-2} M_{pl}^2$ (blue dashed line), $R_0= 5\times10^{-3} M_{pl}^2$ (red dotted-dashed line),  $R_0= 10^{-3} M_{pl}^2$ (brown dotted line), and  $R_0= 10^{-28} M_{pl}^2$ (green dashed line). Notice that for the last two values of $R_0$  no broken phase solutions exist.}
  \label{fig:BR3}
\end{figure}
 
 The backreaction for the case of a  nonminimal coupling to the curvature is illustrated in Fig.  \ref{fig:BR3}, where the left (right) panels correspond to the symmetric (broken) phase solutions. The upper panels illustrate the dependence of the solutions on the coupling to the curvature $\xi_R$, while in the lower panels the coupling $\xi_R$ is fixed and different values for  $R_0$ are considered.
   In particular, from the figure on the bottom left, it can be seen that in the symmetric case, the effect of the quantum corrections may  both  increase or decrease  the value of the  de Sitter spacetime  curvature  $R$ with respect to the classical one, depending on the value of $\Lambda_R$. In  the symmetric phase there are self consistent solutions  for  large values of $\Lambda_R$, while in the broken phase they exist only for  $\Lambda_R$ below a (parameter-dependent) upper bound.  Notice that there is also an upper bound  for $R_0$ bellow which,  under our approximations, no broken phase solution exist  regardless the value of $\Lambda_R$.  On the other hand, one can verify that  the 
 approximation $\mathcal{M}_{ph}^2\ll R$ breaks down for small enough values of $\Lambda_R$  in the  broken phase, and also  in the symmetric case but only when $R_0$ is smaller than a (parameter-dependent) critical value. However, as it can be seen from  the examples considered in the two figures on the left panels, for larger values of $R_0$,  there are symmetric phase solutions where the approximation breaks down for large values of $\Lambda_R$ instead, while remaining valid all the way to $\Lambda_R \to 0$. In these latter cases, we can conclude that there is a divergence of the relative deviation in this limit, which indicates that as $R_{cl}\to 0$, the curvature $R$ goes to a finite positive value. Therefore, for this set of parameters the backreaction
 is crucial to determine the spacetime curvature.

\section{Conclusions}\label{ConcluSect}


In this paper we have considered a self-interacting scalar field with $Z_2$ symmetry in a general curved spacetime.  
In order to include some nonperturbative  quantum effects of the scalar field, we have worked within the Hartree (or Gaussian) approximation to the 2PI EA.

Our first goal  has been to show that in this approximation  the ``consistent renormalization procedure"  described in \cite{Bergesetal} for flat spacetime can be  extended to curved spacetimes
to make finite not only the mean field and gap equations of the matter sector of the theory (which has been  shown in paper I), but also the SEE, which also involve the gravitational sector. That is, we have shown that the same set of counterterms can be used to renormalize the SEE (along with the usual gravitational counterterms that are needed even for free fields). In order to maintain the covariance of the regularized theory, we have  used dimensional regularization.

In Sec. \ref{DSSect}, we have applied  our results to   de Sitter spacetimes. We have considered  the explicit form of the mean value and gap equations, computed  in paper I, together with the SEE for these particular spacetimes, and we have found some self-consistent de Sitter solutions.  The simultaneous solution of the resulting algebraic equations allowed us to discuss the occurrence of spontaneous symmetry breaking and,  at the same time, to assess the effect of quantum fluctuations on the classical metric. 
An important conclusion of our analysis is that the importance of the backreation depends strongly on the value of the curvature at the renormalization point $R_0$. 
We have found both self-consistent solutions where the backreaction is important and solutions where it is not, depending on the values of the parameters.    
In particular, we have found self-consistent de Sitter solutions for vanishing cosmological constant $\Lambda_R=0$, where the quantum effects play a crucial role. 
 
I would be interesting to analyze the spontaneous symmetry breaking and existence of self-consistent solutions beyond the Hartree approximation,  including the setting-sun diagram in the calculation of the effective action. We hope to address these issues in a future work.

\section*{Acknowledgements}
This research was supported in part by ANPCyT, CONICET and UBA. F.D.M and L.G.T. would like to thank ICTP for hospitality during the completion of part of this work.

\end{document}